%% file: reputation.tex
\documentclass[10pt]{article}

\usepackage{times}
\usepackage{verbatim}
\usepackage{subfigure}
\usepackage{color}		
\usepackage{epsfig}
\usepackage{graphicx}
\usepackage{ifpdf}
\usepackage{epstopdf}
\usepackage{amsfonts}
\usepackage{amsbsy}
\usepackage{a4wide}

\makeatletter
\newif\if@restonecol
\makeatother

\usepackage[linesnumbered,ruled]{algorithm2e}

\newif\ifremarks
\remarksfalse 

\ifremarks
\advance\oddsidemargin       -16mm
\advance\evensidemargin    -16mm
\advance\marginparwidth 16mm
\fi

\newcommand{\ms}[1]{\ifmmode%
\mathord{\mathcode`-="702D\it #1\mathcode`\-="2200}\else%
$\mathord{\mathcode`-="702D\it #1\mathcode`\-="2200}$\fi}

{\catcode`\!=8
  \catcode`\Q=3
\long\gdef\given#1{88\fi\Ifbl@nk#1QQQ\empty!}
\long\gdef\blank#1{88\fi\Ifbl@nk#1QQ..!}
\long\gdef\nil#1{\IfN@Ught#1* {#1}!}
\long\gdef\IfN@Ught#1 #2!{\blank{#2}}
\long\gdef\Ifbl@nk#1#2Q#3!{\ifx#3}
}

\def\vo#1#2{\ensuremath{{\sf opinion}\!
                 \if\blank#1\else\left(\ms{#1},\ms{#2}\right)\fi}}

\def\os#1#2{\ensuremath{{\sf os}\!
                 \if\blank#1\else\left(\ms{#1},\ms{#2}\right)\fi}}

\def\e#1#2#3{\ensuremath{{\sf event}\!
                 \if\blank#1\else\left(\ms{#1},\ms{#2},\ms{#3}\right)\fi}}

\def\score#1#2#3#4#5{\ensuremath{{\sf score}\!
                 \if\blank#1\else\left(\ms{#1},\ms{#2},\ms{#3},\ms{#4},\ms{#5}\right)\fi}}

\newcommand{\be}{\begin{equation}}
\newcommand{\bea}{\begin{eqnarray}}
\newcommand{\ee}{\end{equation}}
\newcommand{\eea}{\end{eqnarray}}
\newcommand{\nn}{\nonumber}

\newcommand{\qa}{\alpha}

\newcommand{\qd}{\delta}

\newcommand{\qe}{\varepsilon}

\newcommand{\qz}{\zeta}

\newcommand{\ql}{\lambda}
\newcommand{\qL}{\Lambda}

\newcommand{\qs}{\sigma}

\newcommand{\qt}{\tau}

\newcommand{\prt}{\partial}
\newcommand{\tri}{\triangle}

\newcommand{\one}{{\bf 1}}
\newcommand{\RR}{{\mathbb R}}
\newcommand{\cO}{{\cal O}}

\newcommand{\veca}{\mathbf{a}}
\newcommand{\vece}{\mathbf{e}}
\newcommand{\veceT}{{\bf e}^{\rm T}}
\newcommand{\vecr}{\mathbf{r}}
\newcommand{\vecs}{\mathbf{s}}
\newcommand{\vecv}{\mathbf{v}}
\newcommand{\vecg}{\mathbf{g}}
\newcommand{\vecu}{\mathbf{u}}
\newcommand{\vect}{\mathbf{t}}

\newcommand{\vecvmax}{\mathbf{v}_{\rm max}}
\newcommand{\qlmax}{\ql_{\rm max}}

\newtheorem{example}{Example}
\newtheorem{definition}{Definition}
\newtheorem{theorem}{Theorem}
\newtheorem{lemma}{Lemma}
\newtheorem{corollary}{Corollary}

\begin{document}

\setlength{\parindent}{0mm}

\title{Flow-based reputation: more than just ranking}

\author{Antonino Simone, Boris \v{S}kori\'{c}, Nicola Zannone\\
{\it Eindhoven University of Technology}}

\date{ }

\maketitle

\begin{abstract}
\noindent
{\it
The last years have seen a growing interest in collaborative systems like electronic marketplaces and P2P file sharing systems where people are intended to interact with other people.
Those systems, however, are subject to security and operational risks because of their open and distributed nature.
Reputation systems provide a mechanism to reduce such risks by building trust relationships among entities and identifying malicious entities.
A popular reputation model is the so called flow-based model.
Most existing reputation systems based on such a model provide only a ranking, 
without absolute reputation values; this makes it difficult to determine whether entities are actually trustworthy or untrustworthy.
In addition, those systems ignore a significant part of the available information; 
as a consequence, reputation values may not be accurate.
In this paper, we present a flow-based reputation metric that gives absolute values instead of merely a ranking.
Our metric makes use of all the available information.
We study, both analytically and numerically, the properties of the proposed metric and the effect of attacks on reputation values.
}

\end{abstract}


\input{intro}

\input{rw}

\input{formulation}


\input{properties}

\input{computation}

\input{experiments}

\input{conclusions}

%
%
\bibliographystyle{abbrv}
\bibliography{rep}

\section*{Appendix}

\input{proofs}

\end{document}

%% file: intro.tex
\section{Introduction}

The advent of the Internet has brought new business opportunities and favored the development of collaborative environments.
In particular, the Internet provides the basis for the development of electronic communities where strangers interact with each other and possibly do business.
However, these interactions involve risks. 
For instance, in an eCommerce setting, buyers are vulnerable to risks due to potential incomplete or misleading information provided by sellers \cite{XION-LIU-04-TKDE}.
Similarly, sellers are subject to the risk that the counterparty in a transaction will be unable to honor its financial obligations.
To mitigate those risks, there is the need of a decision support system that is able to determine the trustworthiness of collaborative parties.

Reputation systems are widely considered as `the solution'  to assess trust relationships among users and to identify and isolate malicious users \cite{RKZR-00-CACM}.
Reputation systems are currently adopted in commercial online applications such as P2P file sharing \cite{BitTorrent}, web search \cite{BRIA-PAGE-98-CNISDNS}, electronic marketplaces \cite{amazon,eBay}, and expert systems \cite{allexperts,epinions}.
Reputation is a collective measure of trustworthiness based on aggregated feedback related to past experiences of users.
The basic idea is to let users rate each other and to aggregate ratings about a given user to derive a reputation value.
This value is then used to assist other users in deciding whether to interact with that user in the future \cite{JIB-07-DSS}. 
In the last years, a number of reputation systems have been proposed to aggregate ratings and calculating reputation values; each system is based on a particular theoretical foundation (see \cite{hoffman-09-CSUR,JIB-07-DSS} for a survey). 

The quality of a reputation system is determined by how accurately the computed reputation predicts the future performance of entities \cite{hoffman-09-CSUR}.
This, however, is difficult to achieve because some users can attempt to manipulate their reputation and the reputation of others for their own benefit.
Most existing reputation systems lack the ability to discriminate honest ratings from dishonesty ones.
Therefore, such systems are vulnerable to malicious users who provide unfair ratings \cite{XION-LIU-04-TKDE}.

The issue of discriminating honest from dishonest ratings is usually addressed by reputation systems using the so called flow model \cite{JIB-07-DSS} as the mathematical foundation.
Examples of such systems are EigenTrust \cite{eigentrust}, PageRank \cite{BRIA-PAGE-98-CNISDNS}, SALSA \cite{Lempel2000387}, and PeerTrust \cite{XION-LIU-04-TKDE}.
What makes them appealing is that reputation is computed taking into account the feedback of all the users involved in the system, and the feedback is weighted with respect to the reputation of the user providing the feedback.
Flow models are often based on the theory of Markov chains.
The feedback provided by the users is aggregated and normalized in order to obtain a Markov chain.
Thereby, starting from a vector of initial reputation values, Markov steps are repeatedly applied until a stable state has been reached.

Unfortunately, the current state of affairs regarding this kind of reputation model is not very satisfactory.
First of all, those systems only provide a {\em ranking} of users rather than an absolute reputation value.
Although this can be acceptable in some applications like web search, it is not in others like electronic marketplaces.
For instance, a buyer prefers to do business with an honest seller rather than with the most trustworthy one in a pool of dishonest sellers.
In a scenario where health care providers are willing to use data created by patients \cite{vanDeursen2008159}, 
the quality of the data provided by a patient cannot be assessed by only looking if he is more capable than other patients to do good measurements;
an absolute quality metric is required.
In addition, most of the flow-based systems ignore a significant part of the available information (e.g., negative feedback).
Consequently, the reputation values those systems return may be inaccurate.

To illustrate these points, let us consider an electronic marketplace where users can rate each other after each transaction, as in  eBay \cite{eBay}.
Here, each time Alice has a transaction with another user $j$ (e.g., Bob, Charlie, David), she may rate the transaction as positive, neutral, or negative.
Let us consider scenarios (a) and (b) in Fig.~\ref{fig:example}.
From a reputation metric we would expect that Alice has almost neutral opinion of Bob and Charlie and negative opinion of David in (a), and  positive opinion of Bob and Charlie and neutral opinion of Charlie in (b).
However, if we apply the reputation metric proposed in \cite{eigentrust} to these scenarios, we have that in both (a) and (b) Bob has local trust value\footnote{In \cite{eigentrust} local trust values indicate the opinion that users have of other users based on past experiences. Local trust values are in the range $[0,1]$.} $0.1$, Charlie $0.9$ and David~$0$.
The formulas used to compute these values will be presented in Section~\ref{sec:rw}.
Here, we just want to point out that the metric in \cite{eigentrust} is unable to distinguish between
cases (a) and (b). 
This lack of distinguishing power can be risky for users as it can mislead them in their decision whether to do business with other users.
For instance, the reputation value of Charlie computed using the metric in \cite{eigentrust} in (a) can lead users to think that Charlie is `very' trustworthy while in fact he is not. 
These reputation values only indicate that Charlie is more trustworthy than others (i.e., a ranking without an absolute scale).

Moreover, it is worth noting that in \cite{eigentrust} negative ratings are discarded in order to obtain a Markov chain. 
Consequently, it is not possible to distinguish between users that have (strong) negative reputation and users that have neutral reputation. 
This can be observed by comparing the ratings received by David in scenarios (a) and (b) of Fig.~\ref{fig:example}: although David received a large number of negative ratings and no positive ratings in (a) and an equal number of positive and negative ratings in (b), his reputation value is equal to 0 in both scenarios. 

\begin{figure}[!t]
\centering
\subfigure[]{
{\footnotesize
\begin{tabular}{|l|l|}
\hline
{\bf User} & {\bf Ratings}\\
\hline
Bob & positive 1\\
\cline{2-2}
& neutral 999\\
\cline{2-2}
& negative 0\\
\hline
Charlie & positive 9\\
\cline{2-2}
& neutral 991\\
\cline{2-2}
& negative 0\\
\hline
David & positive 0\\
\cline{2-2}
& neutral 100\\
\cline{2-2}
& negative 900\\
\hline
\end{tabular}
}
\label{fig:subfig1}
}
\subfigure[]{
{\footnotesize
\begin{tabular}{|l|l|}
\hline
{\bf User} & {\bf Ratings}\\
\hline
Bob & positive 100\\
\cline{2-2}
& neutral 900\\
\cline{2-2}
& negative 0\\
\hline
Charlie & positive 900\\
\cline{2-2}
& neutral 100\\
\cline{2-2}
& negative 0\\
\hline
David & positive 200\\
\cline{2-2}
& neutral 600\\
\cline{2-2}
& negative 200\\
\hline
\end{tabular}
}
\label{fig:subfig2}
}
\vspace{-0.2cm}
\caption{\it Example scenarios. 1000 ratings given to Bob, Charlie and David.}
\label{fig:example}
\end{figure}

Last but not least, the design of flow-based reputation models requires including a number of parameters which intend to guarantee the convergence of computations.
However, a comprehensive and exhaustive study of the impact of such parameters on reputation values and how they can be used to protect the systems against attacks has not been conducted yet.

\paragraph{Our contributions}
In this paper, we present a reputation metric that enhances existing flow-based reputation metrics (see \cite{BRIA-PAGE-98-CNISDNS,eigentrust,TRIV-SPIE-ZANN-ETAL-09-PASSAT}) by providing absolute values instead of merely a ranking, and by not discarding any available information.
Computing absolute reputation values makes it possible to quantify the trustworthiness of users and therefore provides a measure to univocally compare reputations values.
This allows us, for instance, to distinguish cases (a) and (b) in Fig.~\ref{fig:example}.
In the design of our reputation metric, we study the effect of self-reference (i.e., a user who gives feedback to himself).
We demonstrate that our construction minimizes such an effect, leading to reputation values that are closer to intuitive expectations.
We formally prove that the proposed reputation metric always has a solution, and that the solution is unique.
We also discuss several methods of solving the reputation equation numerically. 

Our metric depends on a number of parameters: 
a pattern matrix, which stores the (aggregated) feedback received by the system owner from the users about the interactions they had with other users (hereafter, the pattern matrix is also called indirect evidence matrix), a starting reputation vector, which represents the direct information known to the system owner about the trustworthiness of entities in the system, and an interpolation parameter $\qa$, which serves as a weight for direct versus indirect information.
We analytically study the impact of changes in the indirect evidence matrix on reputation values.
This study allows us to analyze how someone can attack the reputation system by providing unfair ratings.
In particular, we analyze self-promoting and slandering attacks\cite{hoffman-09-CSUR} as well as Sybil attacks\cite{DOUC-02-IPTPS}.
To study self-promoting (slandering) attacks, we assume that an attacker can manipulate reputation values by giving positive ratings to users who gave positive ratings to him (negative ratings to the target) and negative ratings to users who gave negative ratings to him (positive ratings to the target).
We study the effect of Sybil attacks by modeling an attacker who subverts the reputation system by first creating a large number of pseudonymous entities, and then using them to influence the reputation value of a target user in a similar way as is done for self-promoting and slandering attacks. 

On the other hand, we assume that the starting reputation vector and the weight parameter are defined by the system owner and cannot be modified by the attacker.
We numerically study the impact of these parameters on reputation values and analyze how they can be used to mitigate the effect of above mentioned attacks. 
The analysis allows us to draw some guidelines for choosing the value of these parameters.
The guidelines are general and apply to reputation metrics that use similar parameters.

In this work we are mainly interested in the study of the mathematical model of reputation systems, 
rather than in the algorithm implementing the mathematical model. 
Therefore, we assume throughout the paper the existence of a central authority which collects all ratings and calculates the reputation of every 
participating user.
This assumption is in line with the approach proposed in \cite{BRIA-PAGE-98-CNISDNS} where a search engine collects information about hyperlinks of several million pages and indexes search results on the basis of such an information.

The paper is structured as follows.
Section~\ref{sec:rw} provides an overview of reputation systems.
Section~\ref{sec:model} presents our metric and
Section~\ref{sec:formal} discusses its formal properties. 
Section~\ref{sec:compute} discusses several methods for computing the reputation vector.
Section~\ref{sec:experiments} evaluates reputations numerically for a number of attack scenarios.
Section~\ref{sec:conclusions} concludes and discusses directions for future work.

%% file: rw.tex
\section{Reputation Systems}
\label{sec:rw}

Reputation systems have been proposed as a mechanism for decision support in open collaborative systems, where entities do not know each other a priori.
Reputation is a collective measure of trustworthiness built from user \emph{experience}. 
A user's experience consists of the events observed by that user.
Events can be, for instance, voiced opinions, that is opinions that are made public \cite{TRIV-SPIE-ZANN-ETAL-09-PASSAT}, downloads \cite{eigentrust}, or transactions \cite{eBay}.
Users can rate the behavior of other users on the basis of their experience.
In particular, \emph{ratings} represent direct judgments of the behavior of users with respect to the perspective of the judging user. 
Those pieces of evidence are aggregated in order to calculate the reputation of users. 
Reputation gives the extent to which the target's behavior is good or bad \cite{Avila-RosasL05}.

In \cite{hoffman-09-CSUR} Hoffman et al.  identify three dimensions of reputation systems: \emph{formulation} (the mathematical model), \emph{calculation} (the algorithm implementing the model and actually computing reputation), and \emph{dissemination} (the mechanism to disseminate the outcome).
Here, we mainly focus on the formulation dimension, and on attacks on the mathematical model.
The formulation of a reputation system includes a number of aspects: \emph{information source}, \emph{information type}, \emph{temporal aspects}, 
and \emph{reputation metrics}.
The source of information can be \emph{subjective}, i.e.\ the rating is based on subjective judgment like in \cite{eBay,TRIV-SPIE-ZANN-ETAL-09-PASSAT}, or \emph{objective}, i.e.\ the rating is determined from formal criteria like in \cite{BRIA-PAGE-98-CNISDNS}.
The advantage of using objective information is that its correctness can be verified by other entities; 
however, sometimes it is difficult to define formal criteria that fully capture entities' opinions. 
At the same time, subjective information makes it difficult to protect the system against unfair rating, which lies at the basis of self-promoting and slandering attacks (see \cite{hoffman-09-CSUR}).
A typical example of these attacks is the so called Sybil attack (see \cite{DOUC-02-IPTPS}), in which different entities or multiple identities held by the same entity collude to promote each other.
Another aspect of information sources is \emph{observability}.
Here it is important whether the information is directly observed by the entity calculating the reputation, or it is obtained second-hand or inferred from direct information.   
We call the reputation value calculated from directly observed information \emph{direct reputation}.\footnote{Direct reputation is also called subjective reputation \cite{Mich-02} or local trust value \cite{eigentrust}.}
Indirect information is widely used in reputation systems to support a notion of transitivity of trust (see \cite{BRIA-PAGE-98-CNISDNS,DKC-07-PODC,JOSA-POPE-05-APCCM,eigentrust}).
Although trust is not always transitive in real life \cite{CHRI-HARB-96-SP}, trust can be transitive under certain semantic constraints \cite{JOSA-POPE-05-APCCM}. In this paper we assume that ratings have the same trust purpose (i.e., the same semantic content) and therefore their aggregation is meaningful.
We also do not distinguish between functional trust (i.e., the ability to make a judgment about a transaction) and referral trust (i.e., the ability to refer to a third party).
As in \cite{eigentrust,XION-LIU-04-TKDE}, we assume that a user trusts the opinion of users with whom he had positive transitions, since users who are honest during transactions are also likely to be honest in reporting their ratings.

The type of information used by a reputation system has a considerable impact on the types of attack to which the system is vulnerable.
Some reputation systems (see \cite{eBay,eigentrust}) allow users to specify ternary ratings (positive, neutral, negative); 
others allow only positive \cite{BRIA-PAGE-98-CNISDNS,TRIV-SPIE-ZANN-ETAL-09-PASSAT} or only negative ratings.
Although systems that only consider positive values are robust to slandering attacks, they are not flexible enough to discriminate between honest and malicious entities. 
Negative reputation systems are particularly vulnerable to whitewashing attacks \cite{hoffman-09-CSUR};
entities who receive a large number of negative ratings can change their identity and re-enter the system with a fresh reputation \cite{LFSC-03-WEPS}.
Therefore, one of our requirements for reputation systems is that entities should not be able to gain an advantage from their newcomer status.
At the same time, newcomers should not be penalized for their status.
Here, the temporal aspects of a reputation system play a fundamental role.
For instance, some systems (see \cite{BRIA-PAGE-98-CNISDNS,JOSA-01-jour,eigentrust,TRIV-SPIE-ZANN-ETAL-09-PASSAT})
do not distinguish between recent and past behavior, 
whereas other systems (e.g., see \cite{Avila-RosasL05,eBay,KRUK-NIEL-SASS-05-CCS}) give more weight to recent behavior.
For instance, in \cite{Avila-RosasL05} reputation values are updated by aggregating the previous reputation value with a factor indicating the proximity of the recent score to the past reputation, i.e.  
$r_{ij}^{(t)} = r_{ij}^{(t-1)} + \mu (\ms{d}_{ij}, r_{ij}^{(t-1)})$, where $\mu$ is a function that determines how fast the reputation value $r_{ij}$ changes after an event with rating $\ms{d}_{ij}$.

A {\em reputation metric} is used to aggregate ratings and compute reputations.
Several computation models have been used: simple summation or average of ratings \cite{amazon,eBay,epinions}, Bayesian systems \cite{JH-07-ARES,TK-09-ENTCS}, beta probability density \cite{vanDeursen2008159}, discrete trust models \cite{CAHI-ET-AL-03-PC}, belief models \cite{ALCA-MAUW-09-FAST,JOSA-01-jour}, 
fuzzy models \cite{BA-09-ECRA,SLK-06-SAC}, and flow models \cite{BRIA-PAGE-98-CNISDNS,eigentrust,Lempel2000387,Walsh2006,TRIV-SPIE-ZANN-ETAL-09-PASSAT}.
Flow models are particularly interesting as they make it possible to compute reputation by transitive iteration through loops and arbitrary chains of entities.
Here, we present the reputation system proposed in \cite{eigentrust} as an example of a flow-based reputation system.
Each time user $i$ has a transaction with another user $j$, she may rate the transaction as positive ($d_{ij}=1$), 
neutral ($d_{ij}=0$), or negative ($d_{ij}=-1$).
The \emph{local trust value} $s_{ij}$ is defined as the sum of the ratings that $i$ has given to $j$,
\be\label{eq:local}
	s_{ij}=\sum_{\rm transactions} d_{ij}.
\ee
This aggregated feedback is then normalized in order to obtain a Markov chain.
Formally, the \emph{normalized local trust value} $a_{ij}$ is defined as follows
\be
\label{eq:normalization}
	a_{ij}=\frac{\max(s_{ij},0)}{\sum_k  \max(s_{ik},0)}.
\ee
Normalized local trust values can be organized in a matrix $[a_{ij}]$ (the so called \emph{pattern matrix}).
In flow models the reputation vector (the vector containing all reputation values) corresponds to the \emph{steady state vector}
of a Markov chain; one starts with 
a vector of initial reputation values and then repeatedly applies the Markov step until a stable state has been reached using the following equation
\be
\label{eq:eigentrust}
	{\bf r}^{(k+1)} = \alpha A^T{\bf r}^{(k)} + (1-\alpha) {\bf p}
\ee
where ${\bf r}$ is the reputation vector, $A$ is the pattern matrix, ${\bf p}$ is a vector of initial reputation values, 
and $\alpha\in[0,1]$ is a damping factor.\footnote{Note that $\qa$ is different from the `$\qa$' in PageRank \cite{BRIA-PAGE-98-CNISDNS}, whose purpose is to modify the matrix.}

Unfortunately, the current state of affairs regarding this kind of reputation model
is not very satisfactory.
First of all, the approach itself has a drawback.
In the Markov chain approach,
reputation values need to be normalized in the sense that they add up to $100\%$ (\ref{eq:normalization}).
The problem is that such reputation values carry {\em relative} information only.
Applying~(\ref{eq:local}) and~(\ref{eq:normalization}) to the two scenarios presented in Fig.~\ref{fig:example}, 
we obtain in both scenarios
that Bob has normalized local trust value equal to $0.1$, Charlie has $0.9$ and David has $0$.
This is good enough for ranking, but when an {\em absolute} measure is required, the Markov chain approach fails.
Actually, one may expect that Bob and Charlie have a similar reputation value in the first scenario; also that the reputation value of Bob in the second scenario is greater than the reputation value of Charlie in the first scenario.
In addition, when entities have a similar reputation value, it is impossible to see whether they are all trustworthy or all untrustworthy.
Suppose a scenario (i) in which Bob and Charlie receive ten positive ratings out of 1000 transactions from Alice and a scenario (ii) in which Bob and Charlie receive $900$ positive ratings out of  1000 transactions.
In principle, Bob and Charlie should have neutral reputation in (i) and strongly positive reputation in (ii).
However, because of the normalization in (\ref{eq:normalization}), from Alice's perspective Bob and Charlie have normalized local trust value equal to $0.5$ in both (i) and (ii).

Furthermore, implementations of flow models ignore a significant part of the available information:
while ratings are positive, negative or neutral, their aggregation ignores the negative
values and maps them to zero.
For instance, EigenTrust \cite{eigentrust} takes the sum $s_{ij}$ of the ratings of all transactions between entities $i$ and $j$, 
and normalizes it with respect to the sum of all the positive ratings given by $i$ (see (\ref{eq:normalization})). 
As a consequence, it is not possible to discriminate between users that have bad reputation and users that have neutral reputation.
Consider, for example, the local trust value of David in the two scenarios of Fig.~\ref{fig:example}: by applying (\ref{eq:local}), we obtain $-900$ in (a) and $0$ in (b). However, after normalizing using (\ref{eq:normalization}), we obtain $0$ in both scenarios.

Finally, the metrics based on Markov chains include parameters which aim to guarantee the convergence of computations and 
to resist malicious coalitions (e.g., the damping factor $\alpha$ and the vector of initial reputation values $s$ in (\ref{eq:eigentrust})). 
Unfortunately, the impact of these parameters on reputation values has not been studied in sufficient detail.

%% file: formulation.tex
\section{Our reputation metric}
\label{sec:model}

\subsection{Reputation model}

Reputation is a collective measure of trustworthiness based on the judgment of a community.
The users in the community can interact with each other and rate the counterpart in the transaction after the completion of the transaction.
The reputation value of a given user is computed by aggregating the ratings that other users in the community gave to that user and reflects the level of trust that they have on the user on the basis of their past experience.
In the remainder of this section, we discuss the rating system, the method for aggregating ratings, and the metric for calculating reputation values from the aggregated ratings.

Ratings are collected by a central authority using a rating system.
We adopt a rating system where ratings are bounded to the corresponding transaction. 
Ratings can be positive, negative, and neutral; we do not impose any restriction on the range of values of ratings. 

The central authority aggregates ratings in order to compute the reputation values of all users involved in the system. 
We assume that aggregated ratings lie in the range $[0,1]$ where $1$ means very good, $0$ very bad, and $\frac{1}{2}$ neutral.
The restriction to $[0,1]$ does not affect the generality of the model: 
values lying in a different interval (and even qualitative values) can easily be mapped to $[0,1]$.
In this way, all the available information (including negative ratings) can be used in the computation of reputation.

A number of factors should be taken into account when ratings are aggregated (see \cite{Avila-RosasL05,TRIV-SPIE-ZANN-ETAL-09-PASSAT,XION-LIU-04-TKDE}):
\begin{itemize}
\item the ratings a user receives from other users,
\item the total number of ratings a user receives from other users,
\item the credibility of the rating source,
\item the size of the transaction, and
\item the time of the transaction.
\end{itemize}
Several aggregation methods based on (some of) these factors have been proposed. In \cite{eBay} ratings are aggregated by summing the positive and negative ratings that the user receives from other users. 
However, it is well known that methods based only on ratings are flawed \cite{Dellarocas2001,XION-LIU-04-TKDE}.
Indeed, a user can increase his reputation by increasing the transaction volume to cover that fact he is cheating at a certain rate.
In particular, the user can build a good reputation in small transactions and then act dishonestly in large transactions \cite{Gan2008}.
To prevent this, an aggregation method should also take into account other factors like the total number of the transactions in which a user is involved and the size of the transaction.
In addition, some existing reputation systems use threshold functions for accurate discrimination between trustworthy and untrustworthy users \cite{TRIV-SPIE-ZANN-ETAL-09-PASSAT}.
In particular, the ratings provided by a user are considered only if the credibility of the user is greater than a certain threshold.
To discriminate between past and recent behavior, some reputation systems update reputation by aggregating the previous reputation with a factor indicating the proximity of the recent rating to the past reputation \cite{Avila-RosasL05}.

The following example presents a simple method for aggregating ratings that incorporates the ratings a user receives from other users, the total number of ratings, and the criticality of the transactions.
Intuitively, the aggregated ratings are defined as the weighted ratio of the sum of positive and negative ratings averaged over the total criticality of transactions.
In the example, we do not consider the credibility of the rating source because this factor is used later in (\ref{selfconsistent}) to calculate reputation values from aggregated ratings.
In (\ref{selfconsistent}), the credibility of a user is given by the reputation of the user.
We refer to \cite{Avila-RosasL05} for an example of a time-sensitive aggregation method.

\begin{example}
Consider the electronic marketplace scenarios of Fig.~\ref{fig:example}.
Let $\mathcal{V}_{xy}$ be the set of transactions between users $x$ and $y$, 
let $q: \mathcal{V}_{xy} \rightarrow \{1,0,-1\}$ be a function that returns the rating given by $y$ to $x$ for the transaction
and $w: \mathcal{V}_{xy} \rightarrow \mathbb{N}$ a function that assigns a criticality value to the transaction.
The aggregated ratings $A_{xy}$ can be computed as the sum of individual ratings weighed with respect to the criticality of the transactions and then mapped into the range $[0,1]$ as follows
\be\label{eq:ex}
	A_{xy}=\frac{1}{2} + \frac{1}{2}\frac{\displaystyle\sum_{\rm v\in\mathcal{V}_{xy}} 
	q(v)w(v)}{\displaystyle\sum_{\rm v\in\mathcal{V}_{xy}} w(v)}.
\ee
If we apply (\ref{eq:ex}) to the scenarios of Fig.~\ref{fig:example} (and assuming that all transactions have the same criticality value), we obtain that the values computed by aggregating the ratings given by Alice to Bob, Charlie, and David in (a) are equal to $0.5005$, $0.5045$, and $0.05$ respectively, 
whereas scenario (b) gives  $0.55$, $0.95$, and $0.5$ respectively.
These values are closer to what one would expect than 
the results of (\ref{eq:normalization}), namely 
$0.1$, $0.9$, and $0$ for Bob, Charlie, and David respectively in \emph{both} 
scenarios. (Here 1 means very good and 0 bad).
\end{example}

The set of all aggregated ratings $A_{xy}$ can be organized in a matrix.
We refer to Table~\ref{tab:repmetric} for the notation used hereafter.

\begin{definition}
\label{def:A}
For $n$ users, the aggregated ratings are contained in an irreducible $n\times n$ matrix $A$,
\begin{itemize}
\item $A_{xy} \in [0,1]\quad$ for $x\neq y$;
\item $A_{xx}= 0$.
\end{itemize}
\end{definition}

$A_{xy}$ represents the aggregated ratings of user $x$ from the perspective of user~$y$.
We impose that self-reference are not included in the aggregation ($A_{xx}=0$ for all $x$).
This choice is motivated in Section~\ref{sec:tmetric}, where we show that a nonzero diagonal 
has undesirable consequences in a simple toy scenario.
In Section~\ref{sec:selfrep} we present numerical results on the effect of self-reference.

To compute reputation, we employ a metric that is an adaptation of the metrics in \cite{eigentrust,TRIV-SPIE-ZANN-ETAL-09-PASSAT}. 
In particular, we adopt the equation proposed in \cite{TRIV-SPIE-ZANN-ETAL-09-PASSAT} (see (\ref{selfconsistent})), which differs from the one proposed in \cite{eigentrust} (see (\ref{eq:eigentrust})) in the moment when the normalization step takes place. 
In \cite{eigentrust} normalization is done once at the beginning in order to obtain a Markov chain using~(\ref{eq:normalization}); 
then, starting from a vector of initial reputation values, Markov steps are repeatedly applied until a stable state has been reached.
Conversely, in \cite{TRIV-SPIE-ZANN-ETAL-09-PASSAT} reputation values are normalized with respect to the sum of all reputation values in the reputation vector ($\sum_z r_z$ in~(\ref{selfconsistent})) at every iteration to guarantee that reputation values stay in the range $[0,1]$.
We differ from the metric proposed in \cite{TRIV-SPIE-ZANN-ETAL-09-PASSAT} in the way the indirect evidence matrix $A$ is defined: in \cite{TRIV-SPIE-ZANN-ETAL-09-PASSAT} $A$ is symmetric (whereas we allow asymmetry), and $A_{xx}=1$ (whereas we set $A_{xx}=0$).

\renewcommand{\arraystretch}{1.4}

\begin{table}[!t]
\centering
{\footnotesize
\begin{tabular}{|l|p{8cm}|}
\hline
{\bf Notation} & {\bf Meaning} 
\\ \hline\hline
$n$ & Number of users.
\\ \hline
$[n]$ & The set $\{1,\cdots,n\}$.
\\ \hline
$\vecr\in[0,1]^{n}$ & Column vector containing all reputations.
\\ \hline
$\vecs\in[0,1]^{n}$ & The `starting' reputation vector. 
\\ \hline
$A_{xy}\in[0,1]$ & Aggregation of ratings of $x$ given by $y$.
\\ \hline
$\qa\in[0,1]$ & Weight of the indirect evidence.
\\ \hline
$\vece$ & The $n$-component column vector $(1,1,\cdots,1)^{\rm T}$.
\\ \hline
$\ell\in[0,n]$ & The `norm' $\veceT\vecr$.
\\ \hline
$\vecv_i$ & The $i$'th eigenvector of $A$.
\\ \hline
$\ql_i$ & The $i$'th eigenvalue of $A$.
\\ \hline
$\qlmax$ & Largest eigenvalue of $A$. 
\\ \hline
$\vecvmax$ & Eigenvector corresponding to $\qlmax$.
\\ \hline
$C$ & The $n\times n$ constant matrix $C=\vece \veceT$. $C_{ij}=1$; $C^k=n^{k-1}C$.
\\ \hline
\end{tabular}
}
\caption{\it Notation}
\label{tab:repmetric}
\end{table}

We consider a system with $n$ users.
The central authority determines the trustworthiness of all users based on his direct experience with them and the aggregated ratings.

\begin{definition}
\label{def:metric}
Let $\vecs\in[0,1]^n$, with $\vecs\neq 0$, be a `starting vector' containing starting values assigned to all users by the central authority.
Let $\qa\in[0,1]$ be a weight parameter for the importance of indirect vs.\ direct evidence.
We define the reputation vector $\vecr\in[0,1]^n$ as a function of $\qa$, $\vecs$ and $A$
by the following implicit equation:
\be
	r_x=(1-\qa)s_x+\qa\sum_{y\in[n]} \frac{r_y}{\ell} A_{xy}
\label{selfconsistent}
\ee
where we have introduced the notation $\ell=\sum_z r_z$.
\end{definition}
Eq.~(\ref{selfconsistent}) can be read as follows.
If the central authority wants to determine the reputation of user $x$,
it first takes into account the direct information that it has about~$x$.
From this it computes $s_x$, the reputation that it would assign to $x$
if it had no further information.
However, it also has the aggregated data in $A$.
It gives weight $1-\qa$ to its `direct' assignment $\vecs$ and weight $\qa$
to the collective result derived from~$A$.
If it did not have any direct information about $x$,
it would compute $r_x$ as $r_x=\sum_y (r_y/\ell)A_{xy}$,
 i.e.\
a weighted average of the reputation values $A_{xy}$ with weights equal to the
normalized reputations of all the users. 
Adding the two contributions, with weights $\qa$ and $1-\qa$,
we end up with (\ref{selfconsistent}),
which has the form of a weighted average over all available information.
Note that (\ref{selfconsistent}) can be expressed in vector notation as
\be
	\vecr=(1-\qa)\vecs +\qa \frac{A\vecr}{\vece^{\rm T}\vecr},
\ee
where $\vece$ stands for the $n$-component column vector $(1,1,\cdots,1)^{\rm T}$.

\subsection{Discussion of self-references} 
\label{sec:tmetric}
The quality of a reputation metric is determined by the accuracy of reputation values.
Here, we provide further motivation for our metric and, in particular, for the choice $A_{xx}=0$.
We demonstrate that the reputation values calculated by our reputation metric are close to the expected values.

The expression for $r_x$ contains a term $\qa (r_x/\ell)A_{xx}$,
the as yet unknown reputation of $x$ multiplied by his 
 `self-rating' $A_{xx}$. 
We briefly investigate the effect of self-reference on our reputation metric.
First we look what happens when the diagonal of $A$ is not set to zero but to $\qz\in[0,1]$.
For large $n$ and random $A$ one does not expect a significant effect, since the diagonal consists of only $n$ elements
out of $n^2$. (See the numerical results in Section~\ref{sec:selfrep}).
We consider the following scenario, which we tailored to make the diagonal stand out:
Everybody agrees that only one user is reasonably trustworthy
(let us call him user~1). Let $\qe\ll 1$ be a small positive constant.
Let $\qs$ be a positive constant of order~1.
We set $A_{xy}=\qe$ for $x\notin \{1,y\}$ and
$A_{1y}=b\in[0,1]$ for all $y\neq 1$.
We set $s_x=\qs\qe$ for $x\neq 1$. 
Because in this scenario all the users except user~1 are treated equally, 
(\ref{selfconsistent}) yields the same reputation for all users $x\neq 1$, which we will denote as
$r_{\rm rest}$. 
\bea
	\!\!\!\!
	A\!=\! \left( \matrix{
	\qz & b   & b    &\cdots & b   \cr
	\qe & \qz & \qe  &\cdots & \qe \cr
	\vdots & \ddots& \ddots   &\ddots & \vdots  \cr
	\vdots &&\ddots& \qz & \qe \cr
	\qe & \cdots & \cdots  & \qe & \qz
	}\right),
	& 
	\vecs\!=\! \left(\matrix{s_1 \cr \qs\qe\cr \vdots\cr\qs\qe}\right),&
	\vecr \!=\! \left(\matrix{r_1 \cr r_{\rm rest}\cr \vdots\cr r_{\rm rest}}\right).
\label{scenario}
\eea
From a good metric we expect that 
user~1 has reputation $(1-\qa)s_1+{\cal O}(\qe)$
and that
$r_{\rm rest}$ is of order $\qe$,
preferably $r_{\rm rest}=(1-\qa)\qs\qe+\qa\qe$.
Substitution of (\ref{scenario}) into (\ref{selfconsistent}) yields, after some algebra,
$r_1=(1-\qa)s_1+\qa\qz+\cO(\qe)$
and $r_{\rm rest}=\qe\frac{(1-\qa)\qs+\qa}{1-\qa\qz/r_1}+\cO(\qe^2)$.
Clearly, our expectations are met only if $\qz=0$.

One could argue that setting the diagonal of $A$ to zero is not enough to remove self-references completely:
in the computation of $r_x$
the normalization factor $\ell=\veceT\vecr$ still contains~$r_x$, 
i.e.\ $r_x$ affects the weights for the computation of $r_x$.
In order to avoid this,
one could define an alternative reputation metric $\vect$ as
\be
	t_x =(1-\qa)s_x+\qa\sum_{y\in[n]\setminus x}
	A_{xy}\frac{t_y}{\sum_{z\in[n]\setminus x}t_z}.
\label{alt}
\ee
For large $n$ and general $A$, the differences between (\ref{alt}) and (\ref{selfconsistent})
are tiny. 
However, substitution of the special scenario (\ref{scenario}) into (\ref{alt})
gives
$t_1=(1-\qa)s_1+\qa b+\cO(\qe)$
and $t_{\rm rest}=(1-\qa)\qs\qe+\qa\qe$.
While $t_{\rm rest}$ is as desired, $t_1$ is not.
There is a significant difference between $t_1$ and the desired outcome
$(1-\qa)s_1+\cO(\qe)$, especially when $b$ is large.
As a special case consider $s_1\ll b$, a situation where the central authority
mistrusts user~1, but all the users trust him.
The authority does not want his result for user~1 to be influenced
heavily by the users, since their reputations are ${\cal O}(\qe)$.

We conclude that the metric $\vecr$ works best when $A_{xx}=0$ is imposed,
and that $\vecr$ is better than the metric $\vect$.
Here `better' means that it more closely matches our expectations of how a metric should behave.

%% file: properties.tex
\section{Formal properties}
\label{sec:formal}

The implicit function (\ref{selfconsistent}) can be shown to
have a number of desirable properties.
In particular, for any choice of $\qa,\vecs,A$ allowed by Definitions~\ref{def:A} and~\ref{def:metric} there always exists a well defined,
unique solution $\vecr\in[0,1]^n$.
This result is fundamental in collaborative systems in which parties rely on the reputation values to make a decision.

In this section, we first introduce some notation and list a number of useful lemmas.
We discuss the trivial solutions for $\qa=0$ and $\qa=1$.
Then, we present a proof
of existence and uniqueness of the solution $\vecr$ for the general case $0<\qa<1$.
Finally, we compute the derivative of $\vecr$ with respect to $A$.
This provides a way to study the sensitivity of the reputation metric to malicious changes in the indirect evidence matrix (Section~\ref{sec:attacks}).

\subsection{Notation and lemmas}
\label{sec:lemmas}

For a vector or a matrix, the notation `$V\geq 0$' means that all the entries are nonnegative.
For other notation we refer to Table~\ref{tab:repmetric}.

\begin{lemma}
\label{lemma:unit}
If  $\vecr$ is a solution of (\ref{selfconsistent})
satisfying $\vecr\geq 0$, then $\vecr\in[0,1]^n$.
\end{lemma}
The proof is given in the Appendix.

\begin{lemma}
\label{lemma:uell}
For given $\qa$, $\vecs$, $A$ 
and a given $\ell\in[0,n]$, 
such that $\det(\ell\one-\qa A)\neq 0$,
there can exist at most one vector $\vecr\in\RR^n$
that satisfies (\ref{selfconsistent}) and $\veceT\vecr=\ell$.
\end{lemma}
{\it Proof:} 
Let $\ell=\veceT\vecr$.
Eq.~(\ref{selfconsistent}) can be rewritten as
\be
	\vecr=\vecu(\ell):=
	(1-\qa)\left[ \one-\frac{\qa}{\ell}A \right]^{-1}\vecs.
\label{uell}
\ee
This fixes the vector $\vecr$ uniquely as a function of the scalar~$\ell$.
\hfill$\square$\\

\noindent
Given a solution $\vecr$, Lemma~\ref{lemma:uell} tells us that
a nontrivial permutation of $\vecr$ cannot be a solution.

\begin{lemma}(Theorem 1.7.3 in Ref.~\cite{BR}) 
\label{lemma:square}
Let $M\geq 0$ be a square matrix. Then $M$ has a positive eigenvalue
$\qlmax$ which is equal to the spectral radius.
There is an eigenvector $\vecvmax\geq 0$ associated with $\qlmax$.
For $x>\qlmax$ it holds that $(x\one-A)^{-1}\geq 0$.
\end{lemma}

\subsection{The special cases $\boldsymbol{\qa=0}$ and $\boldsymbol{\qa=1}$}
\label{sec:alphaspecial}

The case $\qa=0$ trivially yields $\vecr=\vecs$.
The case $\qa=1$ is more interesting.
Eq.~\ref{selfconsistent} reduces to
\be
	A \vecr = \left(\veceT\vecr\right)\vecr.
\label{eigenv}
\ee
This has the form of an eigenvalue equation.
The matrix $A$ has 
eigenvectors $\vecv_i$, and eigenvalues $\ql_i$.
There exist $n$ solutions of (\ref{eigenv}), namely
\be
	\vecr^{(i)} = \ql_i\frac{\vecv_i}{\veceT\vecv_i},
\label{solutiona1}
\ee
i.e.\ proportional to the eigenvectors of~$A$.
However, the Perron-Frobenius theorem for nonnegative irreducible matrices 
(see e.g. Ref.~\cite{BermanPlemmons}, Chapter 2)
tells us that
only one of the eigenvectors gives an acceptable reputation vector: $\vecvmax>0$.
All the other eigenvectors have at least one negative entry.
We are left with a single solution,
\be
	\mbox{At $\qa=1$}:
	\quad\quad\vecr=\qlmax\frac{\vecvmax}{\veceT\vecvmax}
	\quad \mbox{and}\quad \ell=\qlmax.
\label{ralpha1}
\ee

\subsection{The general case $\boldsymbol{0\!<\!\qa\!<\!1}$; Main theorems}
\label{sec:existgeneral}

Multiplying  (\ref{uell}) from the left with $\veceT$
and then multiplying by a suitable constant gives
\bea
\label{fell}
	f(\ell)=1
	& {\rm where}  &
	f(\ell) :=
	(1-\qa)\veceT\left(\ell\one-\qa A\right)^{-1}\vecs.
	\quad\quad
\eea
This equation helps us to prove several important properties of our metric.
First, we demonstrate  that~(\ref{selfconsistent}) has always a well defined, unique solution in the general case $0<\qa<1$.
\begin{theorem}
\label{th:exist}
For $\qa,A,\vecs$ as given in Definitions~\ref{def:A} and~\ref{def:metric}, there exists a reputation vector
$\vecr\in[0,1]^n$ satisfying~(\ref{selfconsistent}).
The solution is of the form $\vecr=\vecu(\ell_*)$ with 
$\vecu$ the function defined in~(\ref{uell}) and 
$\ell_*\in(\qa\qlmax,n]$.
\end{theorem}

\begin{corollary}
\label{corol:limitsqa}
In the limits $\qa\to 0$ and $\qa\to 1$, (\ref{fell})
and~(\ref{uell}) correctly reproduce the reputation vector for the special
cases $\qa=0$ and $\qa=1$.
\end{corollary}

\begin{theorem}
\label{th:unique}
The solution in Theorem~\ref{th:exist} is the {\em only} solution of
(\ref{selfconsistent}) satisfying $\vecr\in[0,1]^n$.
\end{theorem}
The proofs of Theorems~\ref{th:exist} and \ref{th:unique}, and Corollary~\ref{corol:limitsqa} are given in the Appendix.

The quality of a reputation system is determined by how accurately the computed reputation predicts the future performance of entities even when attackers attempt to manipulate reputation values.
The following result allows us to study the effect of unfair ratings by analyzing the sensitivity of reputation values to changes in the indirect evidence matrix.

\begin{theorem}
\label{th:drdA}
For fixed $\qa$ and $\vecs$, a small change in $A$ affects
$\vecr$ as follows: 
\be
	\frac{\prt r_x}{\prt A_{zy}}=\qa \left[\ell\one-\qa A+\frac\qa\ell A\vecr\veceT\right]^{-1}_{xz}r_y.
\label{drdA}
\ee
(Here $[\cdots]^{-1}_{xz}$ stands for element $xz$ of the inverse matrix.)
\end{theorem}
The proof of Theorem~\ref{th:drdA} is given in the Appendix.

Theorem~\ref{th:drdA} gives some direct insight into the effectiveness of attacks.
First, we see that the effect of the attack is proportional to $\qa$.
Furthermore, if some user $y$ wants to attack the reputation of user $x$, 
the most obvious attack is to reduce the matrix
element $A_{xy}$, i.e.\ $(\qd A)_{xy}<0$. We see in (\ref{drdA}) that
the effect is proportional to $r_y$.
Hence, the effectiveness of his attack is proportional to his own reputation.
(Of course this does not come as a surprise, but it is good to see intuition getting confirmed.)
From this we see that it is advantageous for him to improve his own reputation before
attacking other users' reputations.

Finally, from~(\ref{drdA}) we can also read off a less obvious attack strategy.
The attacker $y$ may also indirectly attack $x$ by manipulating $A_{zy}$, where $z$
is some other user. 
The effect of this attack is proportional to the matrix element
$E_{xz}:=[\ell\one-\qa A+\frac\qa\ell A\vecr\veceT]^{-1}_{xz}$.
In practice, user $y$'s attack on $x$ could look as follows.
He computes $E_{xz}$ for all $z$, $z\neq y$.
He picks a number of users $z$ whose $E_{xz}$ have the highest magnitude.
For each of them,
if $E_{xz}<0$, he causes a positive change in $A_{zy}$, otherwise a negative change.
Remark: This reasoning applies for small changes of $\qd A$. 
In the numerical experiments (Section~\ref{sec:attacks})
we take a worst case approach and allow the attacker to make big changes in~$A$.

%% file: computation.tex
\section{Computing reputation}
\label{sec:compute}

From the structure of Lemma~\ref{lemma:uell} and the proof of Theorem~\ref{th:exist}, we can derive a direct method (Fig.~\ref{alg:equation}) for computing $\vecr$ from $\qa$, $\vecs$, and $A$.
This algorithm first solves (\ref{fell}) for $\ell$, obtaining a solution $\ell_*>\qa\qlmax$ (lines~1-3). 
The equation $f(\ell)=1$ is a polynomial equation of degree $n$; this becomes evident if we write $A$ as 
$A=Q \qL Q^{-1}$ (with $\qL$ the diagonal matrix containing the eigenvalues of $A$, and $Q$ the matrix whose columns are the eigenvectors $\vecv_i$) and multiply (\ref{fell}) by $\det(\ell-\qa A)$:
\be
	\prod_{i=1}^n(\ell-\qa\ql_i) = (1-\qa)\sum_{i=1}^n
	(\veceT Q)_i (Q^{-1}\vecs)_i\prod_{j\in[n]\setminus \{i\}}(\ell-\qa\ql_j).
\label{poly}
\ee
The highest order on the left hand side is $\ell^n$, and on the right $\ell^{n-1}$.
The algorithm first completely solves the eigensystem of $A$ (lines~1-2) and then solves (\ref{poly}), looking only for the unique solution $\ell_*>\qa\qlmax$ (line~3).
Finally, it  substitutes that value into (\ref{uell}) (line~4).
Theorem~\ref{th:exist} guarantees that the outcome is a vector in $[0,1]^n$.

\begin{figure}[h]
\centering
\begin{minipage}{5cm}
\fbox{
\begin{minipage}{5cm}
{\footnotesize
\begin{tabbing}
{\scriptsize 1}\ \ \= $\{\ql_i\}= {\sf Eigenvalues}(A)$\\
{\scriptsize 2} \> $Q= {\sf Eigenvectors}(A)$\\
{\scriptsize 3} \> Find $\ell_*>\qa\qlmax$ that solves (\ref{poly})\\
{\scriptsize 4} \> $\vecr=(1-\qa)\left[ \one-\frac{\qa}{\ell_*}A \right]^{-1}\vecs$
\end{tabbing}  
}
\end{minipage}
}
\caption{Direct method}
\label{alg:equation}
\end{minipage}
\hspace{0.5cm}
\begin{minipage}{5cm}
\fbox{
\begin{minipage}{5cm}
{\footnotesize
\begin{tabbing}
{\scriptsize 1}\ \ \= $\vecr^{(0)} = \vecs$\\
{\scriptsize 2} \> {\bf re}\={\bf peat}\\
{\scriptsize 3} \> \> $\vecr^{(k+1)} = \frac{A\vecr^{(k)}}{\sum_z r^{(k)}_z}$\\
{\scriptsize 4} \> \> $\vecr^{(k+1)} = (1-\qa)\vecs +\qa\vecr^{(k+1)}$\\
{\scriptsize 5} \> \> $\ms{diff}= \|\vecr^{(k+1)}-\vecr^{(k)}\|_1$\\
{\scriptsize 6} \> {\bf until} $\ms{diff}<\delta$
\end{tabbing}  
} 
\end{minipage}
}
\caption{Iterative method}
\label{alg:iterative}
\end{minipage}

\end{figure}

An iterative method for solving (\ref{selfconsistent}) is presented in Fig.~\ref{alg:iterative}.
This algorithm first computes reputation as the weighted average of reputation values in $A$ (line~3).
Then, it calculates the average over direct and indirect evidence using $\qa,1-\qa$ as weights (line~4).
The algorithm repeatedly computes the reputation vector until it converges, that is, the difference between the new state $\vecr^{(k+1)}$ and the previous one $\vecr^{(k)}$ is less than a certain threshold (lines~5-6).
Notice that the termination condition corresponds to
\be
	\left\| \quad
	(1-\qa)\vecs +\qa \frac{A\vecr^{(k)}}{\mbox{$\sum_z$} r^{(k)}_z}-\vecr^{(k)}
	\quad \right\|_1<\delta.
\label{stopcrit}
\ee
In Section~\ref{sec:compare} we show numerically that the two algorithms find the same solution.

%% file: experiments.tex
\section{Numerical experiments}
\label{sec:experiments}

In this section, we assess the performance of our metric for different choice  of the parameters $\qa$ and $\vecs$ and discuss how these parameters can be used to mitigate the impact of attacks on the reputation system.
In particular, we first discuss our choice for the $A$-matrix.
We compare the performance of the algorithms for computing the reputation vector $\vecr$ presented in Section~\ref{sec:compute}.
Then, we investigate the effect of $\qa$ and $\vecs$ as well as the effect of self-ratings
on $\vecr$.
Finally, we discuss attacks and their effectiveness.

\subsection{Generation of the matrix $\boldsymbol{A}$}
\label{sec:genA}
To study our metric, we simulate a characteristic marketplace scenario.
Our scenario consists of a number of users who can interact with each other and rate the party with whom they interact after a transaction.
In our experiments, we also investigate the robustness of the metric against different threat models which describe typical attacker behavior.
Threat models will be described in Section~\ref{sec:attacks}.

In order to simulate a realistic scenario,
we generated random $A$-matrices as follows.
\begin{enumerate}
\item
All non-diagonal elements of $A$ are initialized to~$\frac{1}{2}$.
This is the `neutral' value for users who have not yet interacted with each other.
\item
For each user $i$, a value $\qt_i\in[0,1]$ is drawn from a triangular probability distribution
$\qs(\qt)$ that has $\qs(0)=0$, $\qs(1)=0$, and a peak $\qs(\qt_{\rm max})=2$.
The number $\qt_i$ serves as the `intrinsic' trustworthiness of user~$i$.
We have a group of experiments with varying $\qt_{\rm max}$ to show its effect and, otherwise, $\qt_{\rm max}$ is set to $0.6$ as the representative value.
\item
We fix a number $f\in(0,1)$, the `filling fraction'. We randomly generate
$f(n^2-n)$ user pairs $(x_a,y_a)$, with $x_a\neq y_a$.
These pairs represent past interactions between the selected users, where $y_a$ judged~$x_a$.
We set $f=0.3$.
\item
For each of the pairs $(x,y)$
the matrix element $A_{xy}$ is assigned a random value
uniformly drawn from the interval
$[\max\{\qt_x-0.1 , 0\}, \min\{\qt_x+0.1,1\}]$.
This step simulates the fact that the judgment of $x$ by $y$ is mostly determined by
the intrinsic trustworthiness $\qt_x$, while allowing for some noise.
\end{enumerate}
We consider this set-up acceptably realistic for the following reasons. 
First, for large $n$ it is unlikely that every user has interactions with everybody else. 
Only a fraction $f<1$ of the matrix gets `filled'. 
Second, the direct opinion about a user is the result of interactions with him. 
Someone's opinion about $x$ depends mainly on the behavior of $x$ (whose intrinsic trustworthiness is modeled as $\qt_x$), 
and also on other circumstances, which we model as small-amplitude random noise.
Our choice of a triangle-shaped probability distribution for $\qt$
is motivated by the wish to keep the model as simple as possible while still containing the
necessary ingredients.

\subsection{Comparison of computation methods}
\label{sec:compare}

We implemented the algorithms presented in Figs.~\ref{alg:equation} and~\ref{alg:iterative} in Wolfram Mathematica~7.0.
It turns out that the iterative method (Fig.~\ref{alg:iterative}) is faster than the direct method (Fig.~\ref{alg:equation})
at the same level of accuracy.
This is hardly surprising, since the heaviest operations in the iterative method
are the repeated matrix-times-vector multiplications (order $n^2$ times the number
of iterations), while the direct method involves solving the whole eigensystem of an $n\times n$ matrix.
We did a number of experiments where we solved $\vecr$ with the direct method,
using Mathematica's default machine precision.
This gave 
\be
	\left| 
	\; r_i - \left[ (1-\qa)\vecs+\qa \frac{A\vecr}{\veceT\vecr} \right]_i 
	\;\right|< 10^{-15}.
\label{accalg2}
\ee
We did the same experiments (same $A$, $\qa$, $\vecs$) with Alg.~2,
with $\qd=n\cdot 10^{-15}$.
This $\qd$ is tailored to yield the same accuracy as (\ref{accalg2}),
as can be seen by comparing (\ref{accalg2}) to (\ref{stopcrit}).

\begin{figure}[!t]
\centering
\includegraphics[width=58mm]{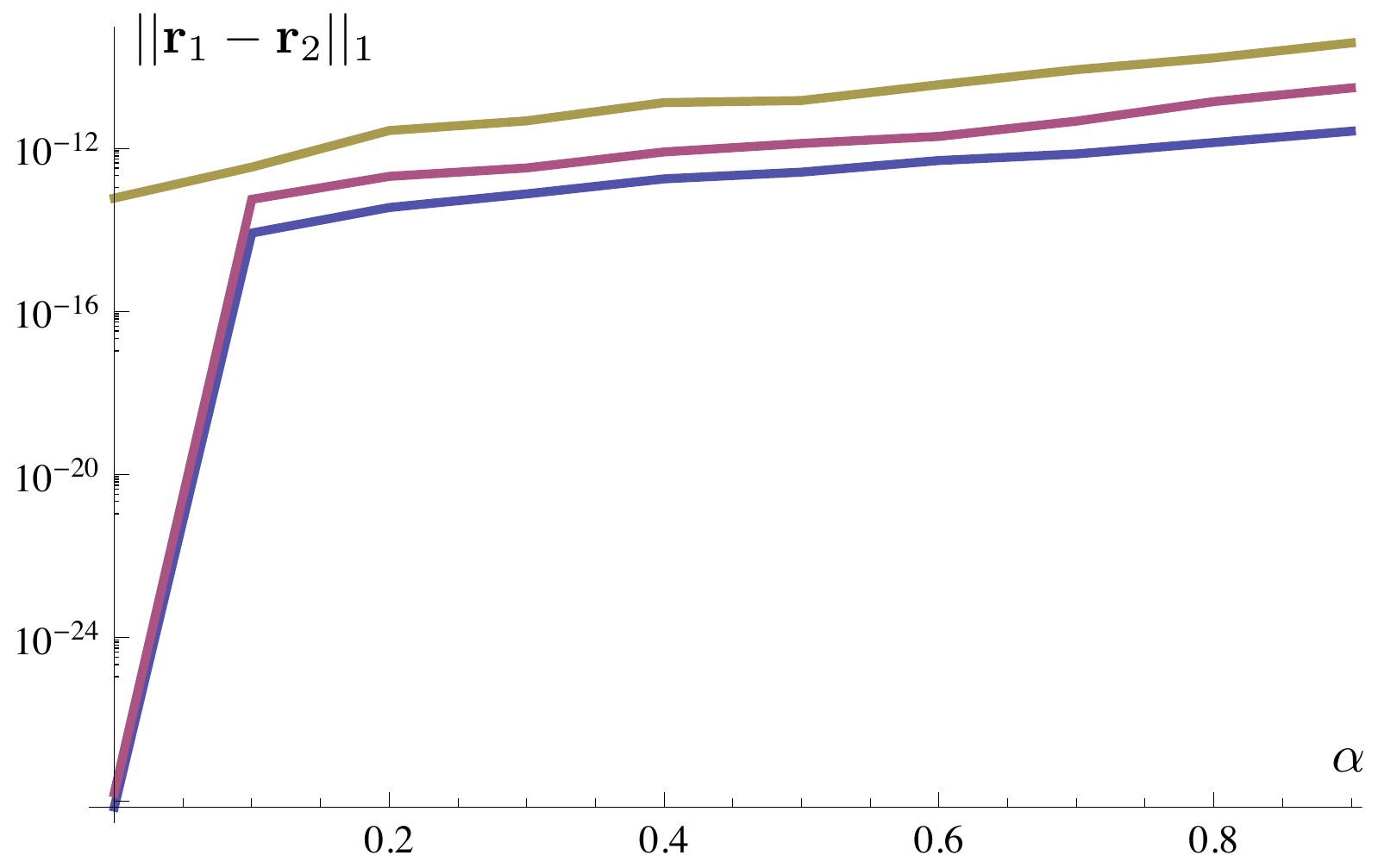}
\caption{\it Logarithmic plot of the difference $|| \vecr_1-\vecr_2 ||_1$ 
between the two algorithms,
averaged over 20 experiments, as a function of~$\qa$.
From bottom to top $n=50, 100, 200$.}
\label{fig:diff}
\end{figure}

For $n$ we took the values 50, 100, and 200.
The number of required iterations is then typically 12 or less,
and decreases with growing $n$. 
For every $n$ and $\qa$ we took 20 different $A$-matrices.
Fig.~\ref{fig:diff} shows the distance $|| \vecr_1-\vecr_2 ||_1$
averaged over these 20 experiments, where $\vecr_1$, $\vecr_2$ are the
solutions found by the direct and iterative method respectively.
Clearly $\vecr_1$ and $\vecr_2$ are almost identical.

The results presented hereafter were obtained using the iterative method.

\subsection{Impact of the parameters}
\label{sec:numparams}

The objectives of this set of experiments is to evaluate the impact of parameter $\qa$ and initial reputation vector $\vecs$ on the reputation vector $\vecr$.

\begin{figure}[!h]
\centering
  
\subfigure[]{
\includegraphics[width=60mm]{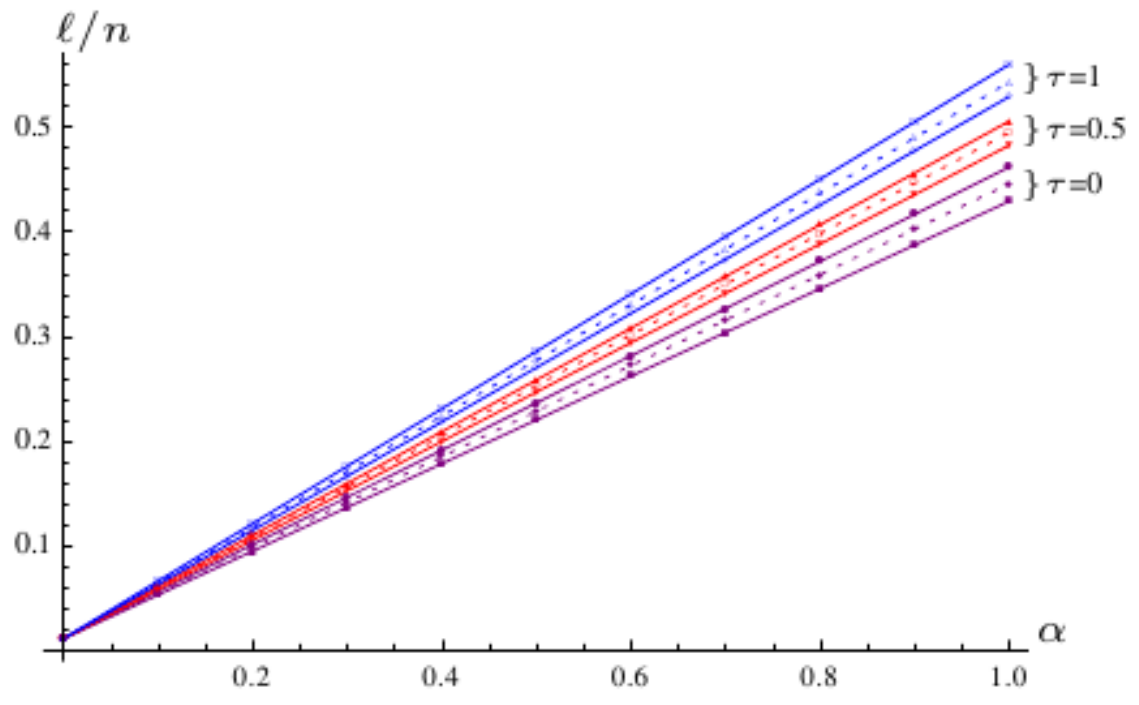}
	\label{fig:qa_l}
}
\hspace*{0.5cm}
\subfigure[]{
    \includegraphics[width=56mm]{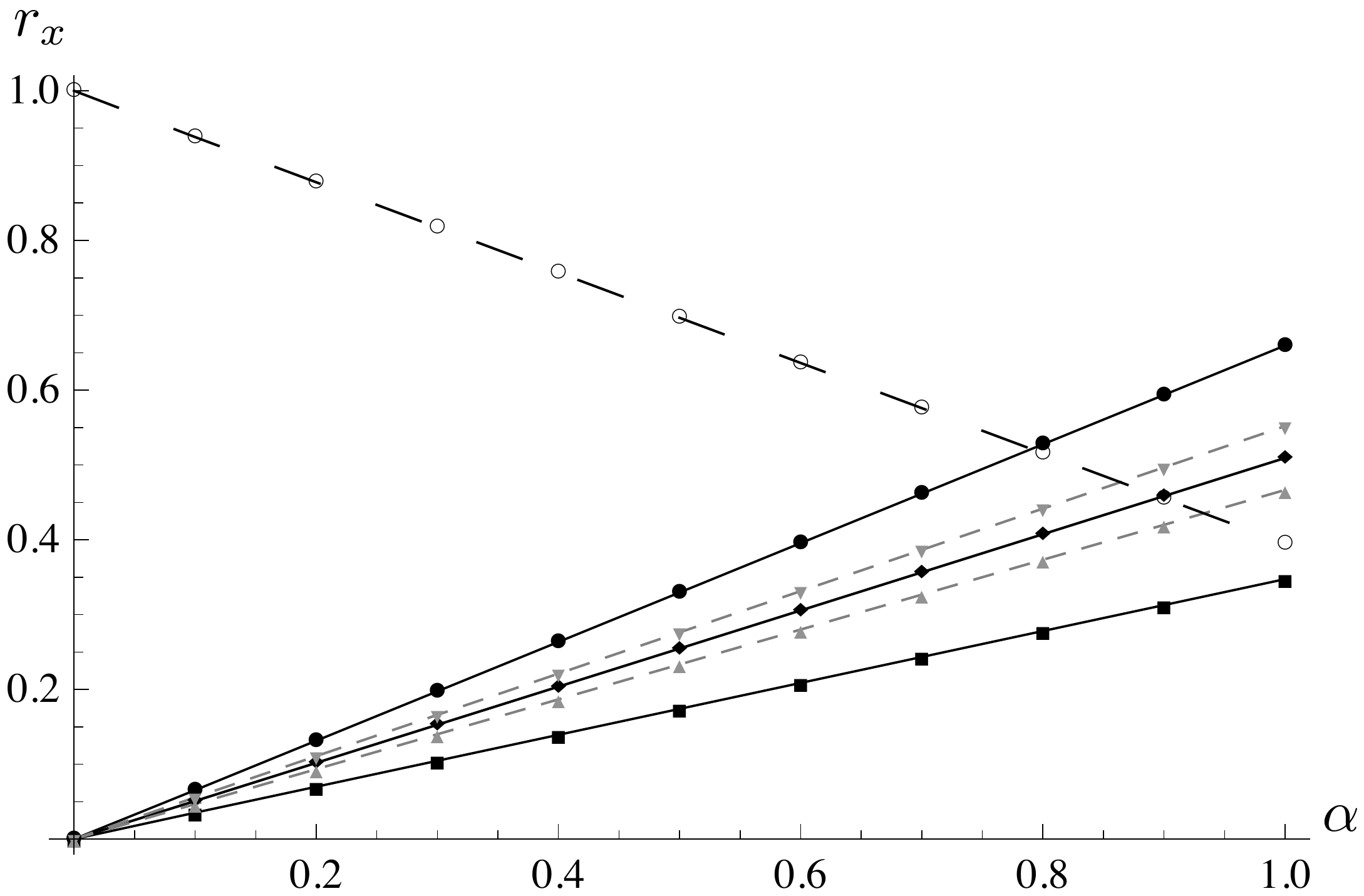}
	\label{fig:qa_r}
	}
\caption{\it Dependence of the reputation on $\qa$. $n=500$ and $\vecs=(1,0,\cdots,0)^{\rm T}$.
{\bf (a)} For every $\qa$, 50 random $A$ were taken, and only the min., max. and average $\ell/n$ are plotted.
{\bf (b)} One fixed random $A$. The downward curve is $r_1$.
The other plotted components are those with the minimum, maximum and median 
reputation at $\qa=1$, plus two more in between.}
\label{fig:qa}
\end{figure}

\paragraph{The parameter $\boldsymbol\qa$.}
Figs.~\ref{fig:qa_l} and \ref{fig:qa_r} show $\ell$ and selected components of $\vecr$ as a function of $\qa$.
The linearity in these graphs is surprising, since
we know from (\ref{selfconsistent}) that $\vecr$ is {\em not} strictly linear in $\qa$. 
(Close inspection of Fig.~\ref{fig:qa_r}
indeed shows that the data do not precisely lie on straight lines.)
Yet $\vecr$ is quite close to a linear interpolation between the $\qa=0$ and $\qa=1$
solutions,
\be
	\vecr \approx (1-\qa)\vecs +\qa \frac{\qlmax\vecvmax}{\veceT \vecvmax}.
\label{linear}
\ee
This result is independent from the choice of $\qt_{max}$ as shown in Fig.~\ref{fig:qa_l}.
As expected, $\qt_{max}$ has an impact on the average reputation of peers within the system ($\ell /n$). 
Fig.~\ref{fig:qa_r} demonstrates how a pre-trusted user (a user who has initial reputation equal to $1$) can lose his leading position
when $\qa$ increases, as gradually more weight is given to $A$ than to $\vecs$.

As we discussed earlier, $\qa$ serves as weight for direct versus indirect information.
Accordingly, the system owner should choose the value of $\qa$ on the basis of his confidence in the information he initially has.
Suppose, for instance, that he is confident that a user $x$ is trustworthy, but his reputation $r_x$ 
turns out below average.
This may arouse suspicion that some malicious user is attempting to subvert the system.
In this case, he should select a low value of $\qa$ to reduce the influence of the information provided by users on the computation of reputation values.
At the same time, setting $\qa$ to $0$ would make it impossible to capture the dynamics of the actual user behavior.
The study of the behavior of the components of $\vecr$ (Fig.~\ref{fig:qa_r}) can assist the system owner to select $\qa$ in such a way that $x$ keeps his high ranking, while information provided by users is still taken into account.

\begin{figure}[!t]
\centering
\subfigure[]{
\includegraphics[width=45mm]{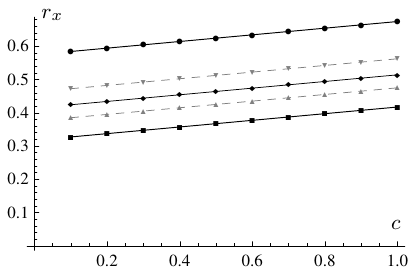}
\label{fig:sdep_a02}
}
\subfigure[]{
	\includegraphics[width=45mm]{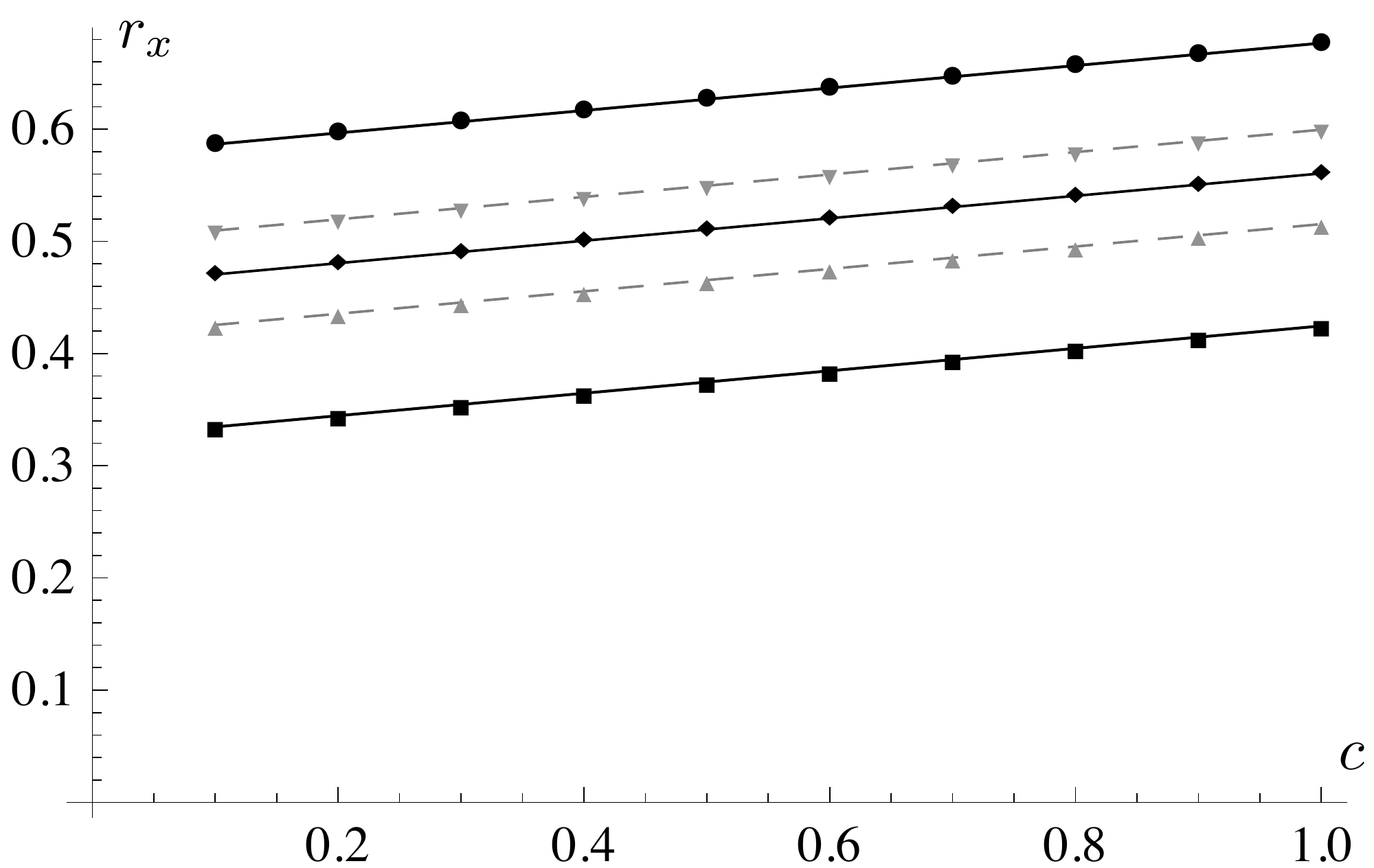}
	\label{fig:sdep_a05}
	}
\subfigure[]{
\includegraphics[width=45mm]{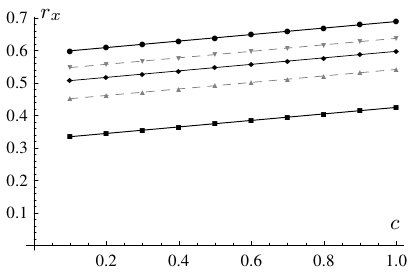}
\label{fig:sdep_a09}
}
\caption{\it Dependence of the reputation on $\vecs$. $n=1000$, $\qa=0.9$, one (typical) fixed random~$A$, $\vecs=c\,\vece$. 
{\bf (a)} $\qt_{max}=0.2$;
{\bf (b)} $\qt_{max}=0.6$;
{\bf (c)} $\qt_{max}=0.9$.}
\label{fig:sdep1}
\end{figure}

\begin{figure}[!t]
\centering
	\includegraphics[width=58mm]{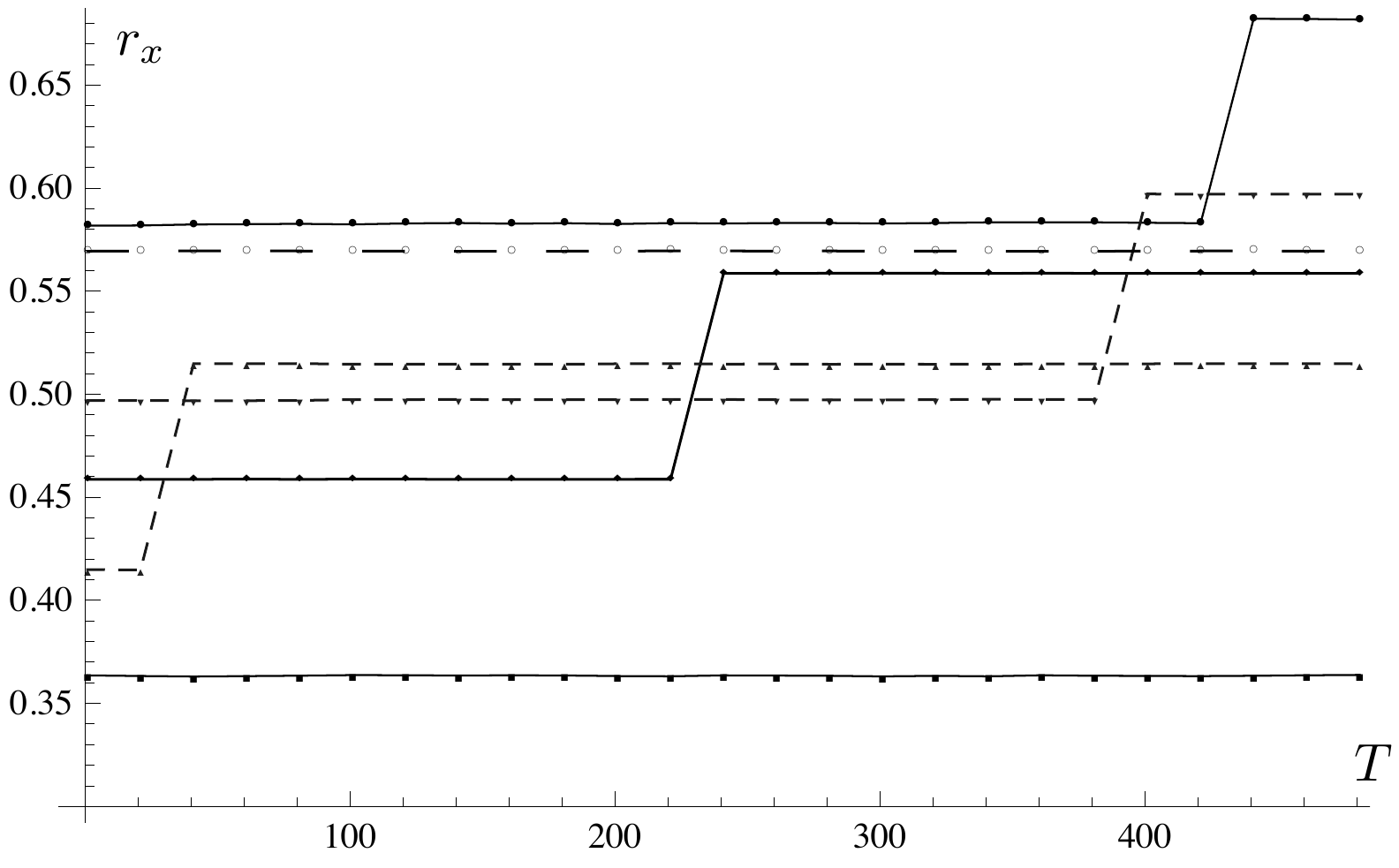}
\caption{\it Dependence of the reputation on $\vecs$. $n=1000$, $\qa=0.9$, one (typical) fixed random~$A$, $\vecs=(1,\cdots,1,0,\cdots,0)$, with $T$ pre-trusted users.}
\label{fig:sdep2}
\end{figure}

\paragraph{The parameter $\vecs$.}
We studied the dependence on $\vecs$ in two ways:
(i) We set $\vecs=c \vece$, with $c\in[0,1]$, i.e., the initial reputation given by the system owner is equally distributed among all users (Fig.~\ref{fig:sdep1}). We repeated the experiments for different $\qt_{\rm max}$ values.
(ii) We set $\vecs=(1,1,\cdots,1,0,\cdots,0)^{\rm T}$, varying the number $T$ of 1s in the vector
(i.e., the number of pre-trusted users).
Fig.~\ref{fig:sdep2} shows several components of $\vecr$ for these two cases.
For $\vecs=c\,\vece$, 
the linear behavior of $\vecr$ as a function of $c$ 
is hardly surprising, in view of the approximation (\ref{linear}) which is linear in $\vecs$. 
Fig.~\ref{fig:sdep1} also shows that the average reputation of peers within the network increases with the increase of $\qt_{\rm max}$.
Case (ii) shows jumps as a function of $T$, and even the ranking changes occasionally.
This can also be understood from (\ref{linear}). 
When an extra user $x$ is included in the pre-trusted set,
the main effect is a jump in $r_x$ of size $\approx 1-\qa$, with only minor changes to the other reputations. 
This result demonstrates that the effect of selecting pre-trusted users wrongly can be mitigated by increasing~$\qa$.

In summary, the starting vector $\vecs$ has a clear effect on the reputations $\vecr$, 
which is well described by the linear approximation (\ref{linear}).
In particular, $\vecs$ makes $\vecr$ less sensitive to changes in $A$.
The pre-trust that the central authority puts in users is carried over 
(multiplied by a factor $1-\qa$) into the reputation vector $\vecr$.
Some guidelines for choosing $\vecs$ are given in Section~\ref{sec:attacks}.

\subsection{Effect of self-references}
\label{sec:selfrep}

As discussed in Section~\ref{sec:model}, we set $A$'s diagonal  to $0$ 
to minimize the effect of self-references. 
In this section we study what happens when the diagonal is set to~1 (the strongest possible departure from $A_{xx}=0$).

In our experiments, we create an $A$ matrix for different $n$ (from 10 to 500) and calculate reputation when the values on the diagonal are $0$ (obtaining $\vecr_0$, with norm $\ell_0=\veceT\vecr_0$) and when they are $1$ 
(obtaining $\vecr_1$, with norm $\ell_1=\veceT\vecr_1$). 
We use the relative change $\tri\ell/\ell_0=(\ell_1-\ell_0)/\ell_0$ as a measure of the influence of self-references.
We performed 20 experiments for each $n$ and for $\qa=0.1, 0.5, 0.9$; for each set of experiments, we determined the average change. 

\begin{figure}[!t]
\centering
\includegraphics[width=59mm]{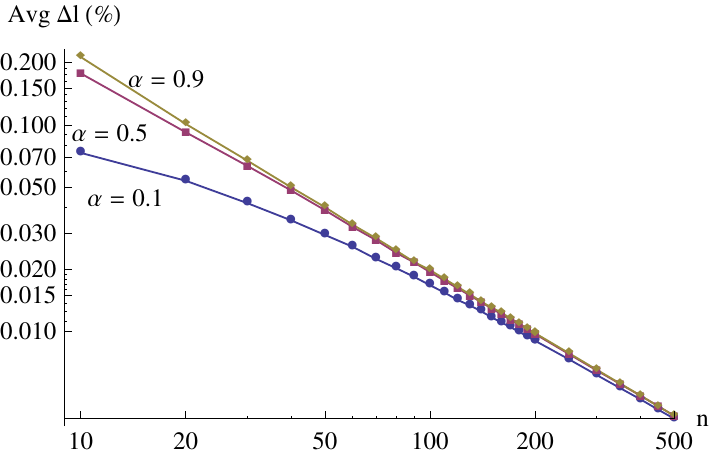}
\caption{\it Effect of self-rating. The average change $\tri\ell/\ell_0$ as a function of $n$ 
(for various values of $\qa$) plotted on a log-log scale. The slope $-1$ indicates that $\tri\ell/\ell_0 \propto n^{-1}$
for large~$n$.}
\label{fig:selfrep}
\end{figure}

Fig.~\ref{fig:selfrep} shows the average percentage change in $\ell$ as a function of $n$.
We observe that the magnitude of the change is inversely proportional to $n$, and hence becomes negligible for large $n$.
The effect of self-references is non-negligible at small $n$.
For $n$ between $10$ and $200$, it varies between 20\% and 1\%.
Furthermore, 
we observe that $\vecr_1=(1+\frac\qa{\ell_0})\vecr_0$.
This proportionality is explained in the Appendix. 
As a consequence, $\ell_1-\ell_0=\qa$.
Notice that this result is constant, independent of $n$.
In contrast, $\ell_0$ depends on $n$ linearly in most cases.
(For specially crafted $A$ and $\vecs$, such as the scenario in Section~\ref{sec:tmetric}, 
$\ell_0$ may be independent of~$n$.)
This explains the inverse-$n$ proportionality of the percentage change $\tri\ell/\ell_0$.

\subsection{Robustness Against Attacks}
\label{sec:attacks}

In our model,
we assume that malicious users can compromise the integrity of  $\vecr$ only by manipulating $A$ (i.e., by providing unfair ratings). 
They can influence neither $\qa$ nor $\vecs$.
Attacks on the computation process and dissemination
are out of scope in this paper.
Whitewashing attacks are not critical, since new users get neutral entries
in $A$.
We assume that $A$ is publicly known.
How strongly $A$ can be manipulated
depends on the actual feedback aggregation method.
This can be a slow and/or costly process, e.g. if
it involves feedback on transactions.
In Section~\ref{sec:existgeneral} we described the effect of a small change in $A$ and discussed how an attacker can exploit such changes to affect the computation of reputation.
Here, we present some threat models that are inspired on such a malicious behavior:
\begin{description}
\item[Self-promotion:]
The attacker's goal is to improve his own reputation.
He can do this by giving 
(i) positive feedback to
users who have given him positive ratings, and (ii) negative feedback to
users who have given him negative ratings.
\item[Slandering:]
The goal is to ruin the reputation of a target $x$.
The options are giving (i) negative feedback to $x$, (ii) positive feedback to users
who have given negative ratings to $x$, and (iii) negative feedback to users
who have given positive ratings to~$x$.

\newpage

\item[Sybil attacks:]
An attacker creates new accounts.
These give positive feedback to him and to each other in order to improve
their reputation and ability to influence $\vecr$.
Then, a slandering attack can be launched with the help of the new accounts.
\end{description}

Let $t$ be the number of attackers.
Without loss of generality we can group the attackers
together and let them lie in $\{n-t+1,\cdots,n\}$.
Then, $A$ is of the form
\be
	A=\left(\begin{array}{c|c}
	\mbox{honest judged} & \mbox{honest judged} \\
	\mbox{by honest} & \mbox{by attackers} \\
	\hline
	\mbox{attackers judged } & \mbox{ attackers judged} \\
	\mbox{by honest} & \mbox{by attackers}
	\end{array}\right)
\ee
Only the right hand part of the matrix can be influenced by the attackers.
A Sybil attack enlarges $A$ by adding rows at the bottom and columns at the right.

\begin{figure}[h]
\centering
\subfigure[]{
	\includegraphics[width=5.8cm]{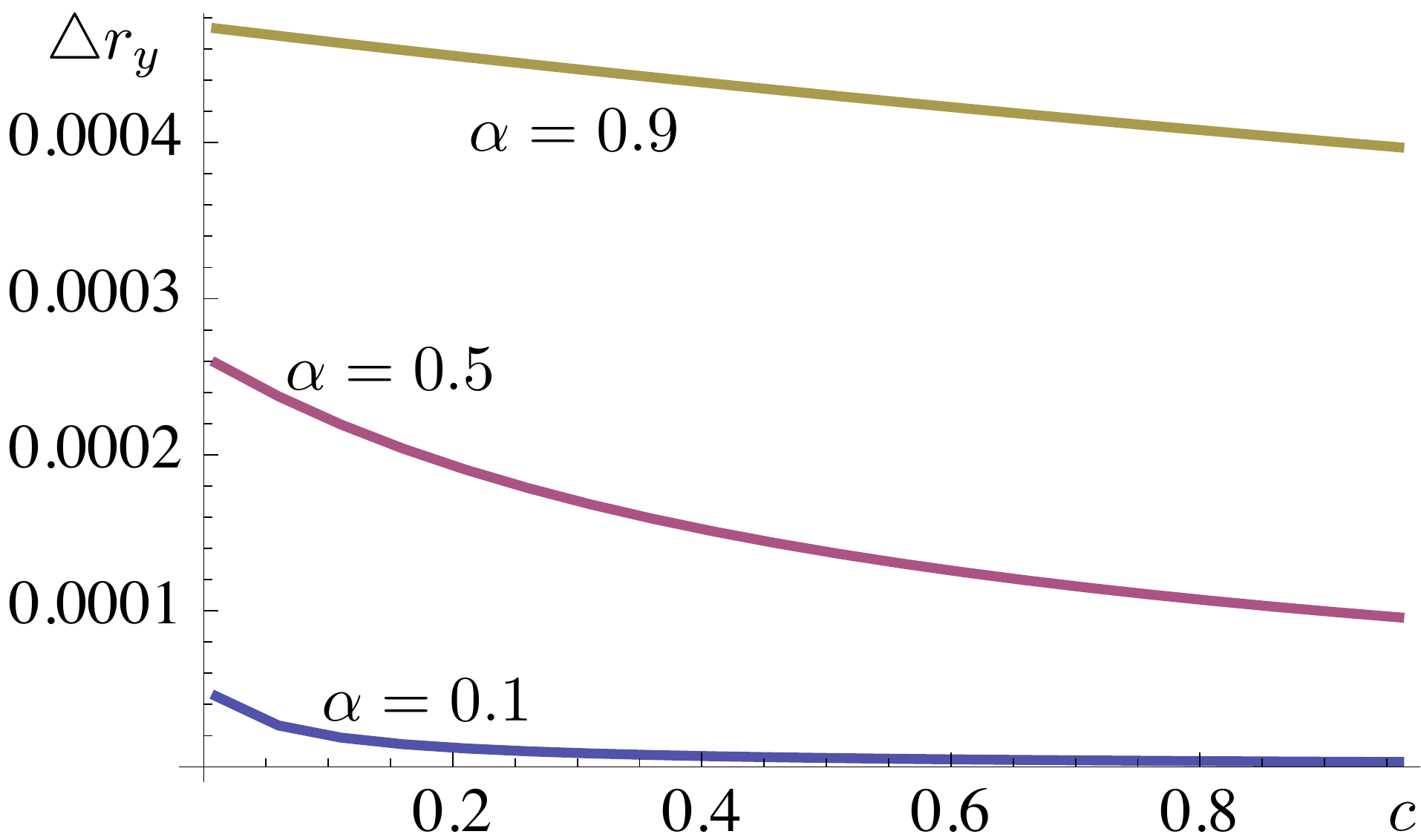}
	\label{fig:selfprom1}
	}
\hskip10mm
\subfigure[]{
	\includegraphics[width=5.8cm]{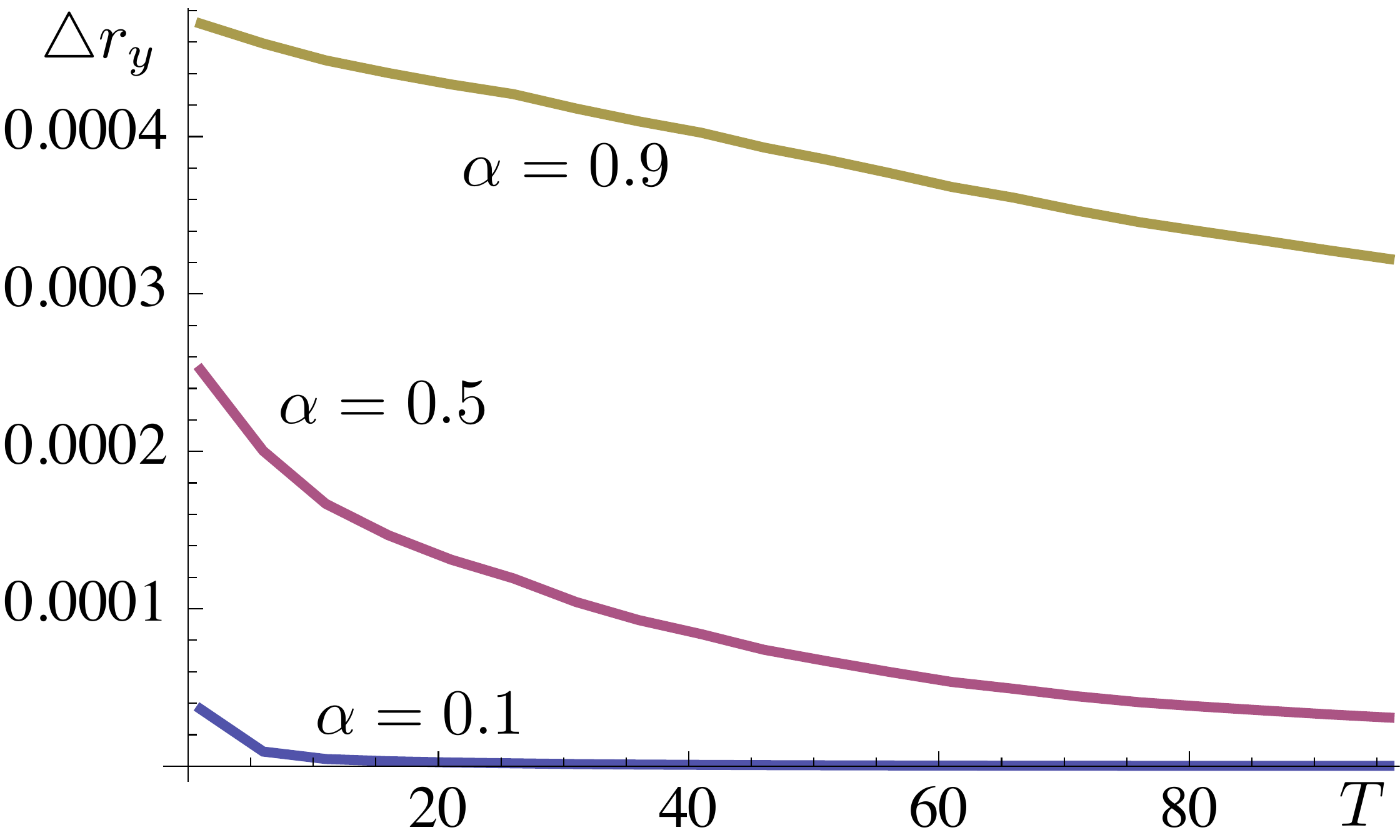}
	\label{fig:selfprom2}
	}
\caption{\it Effect of self-promotion attacks. $n=200$, single fixed~$A$.
The difference in the attacker's reputation is shown (a) as a function of $c$, for
$\vecs=c\vece$; (b) for $\vecs=(1,\cdots,1,0\cdots,0)$ as a function of the number $T$ of pre-trusted users.
(The attacker is not one of the pre-trusted users.)}
\label{fig:selfprom}
\end{figure}

\paragraph{Self-promotion experiments.}
The objective of this set of experiments is to evaluate the effectiveness and robustness of the reputation metric against self-promotion attacks.
Consider one attacker~$y$. He modifies $A_{xy}$ to 1 if $A_{yx}>0.5$
and to $0$ if $A_{yx}<0.5$.
(He tries to boost the reputation of those that have a high opinion of him,
and to reduce the reputation of the rest.)
The effect $\tri r_y$ of such an attack is shown in Fig.~\ref{fig:selfprom}.
The plotted data are calculated for one random~$A$,
but we have performed many such experiments; the presented results are typical
for the whole ensemble of random matrices. We chose an attacker $y$ with $r_y<0.5$,
i.e. a user with less than neutral reputation who actually needs the attack.\footnote{
An attacker with $r_y>0.5$ has a bit more effect (\ref{drdA}), unless $r_y$ is close to~1,
where no more improvement is possible. 
}

Clearly the attacker has little effect; his opinion is only one of many.
As expected,
$\tri r_y$ grows with~$\qa$, since any change in $A$ gets weight $\qa$ in the computation of the reputation.
We also see that the choice of $\vecs$ has a nontrivial impact. 
In particular, the larger the (total) reputation that the system owner initially gives to users, the smaller the effect of the attack.
Both Figs.~\ref{fig:selfprom1} and~\ref{fig:selfprom2} show a nonlinear dependence
of $\tri r_y$ on the components of $\vecs$.
Note that the attack strength for $c\to 1$ is not the same as for 
$T\to n-1$. 
In particular, $s_y$ is not the same in these cases: in the first case the attacker has initial reputation $s_y =c$, whereas in the second case $s_y =0$.
This turns out to have a noticeable effect. 

In summary, the most effective countermeasure for mitigating self-promotion attacks is to decrease $\qa$.
Another strategy would be to enlarge the set of pre-trusted users.
It is worth noting that this result contradicts the suggestion given in \cite{eigentrust} to choose a very small number of pre-trusted users.
However, if the attacker is included in the set of pre-trusted users, the power of the countermeasure is reduced.

\begin{figure}[!t]
\centering
\subfigure[]{
	\includegraphics[width=5.8cm]{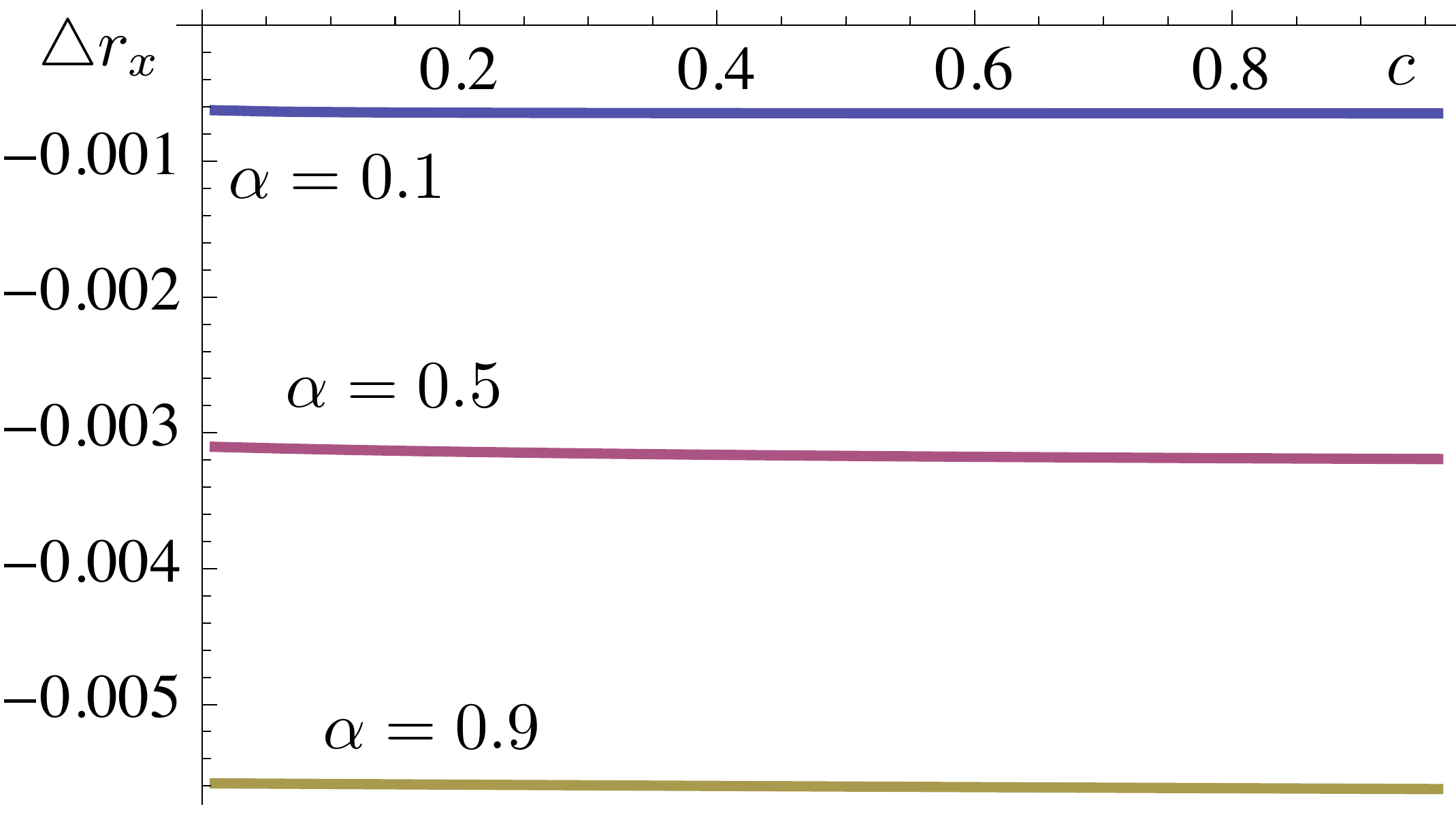}
	\label{fig:slander1}
	}
\hskip10mm
\subfigure[]{
	\includegraphics[width=5.8cm]{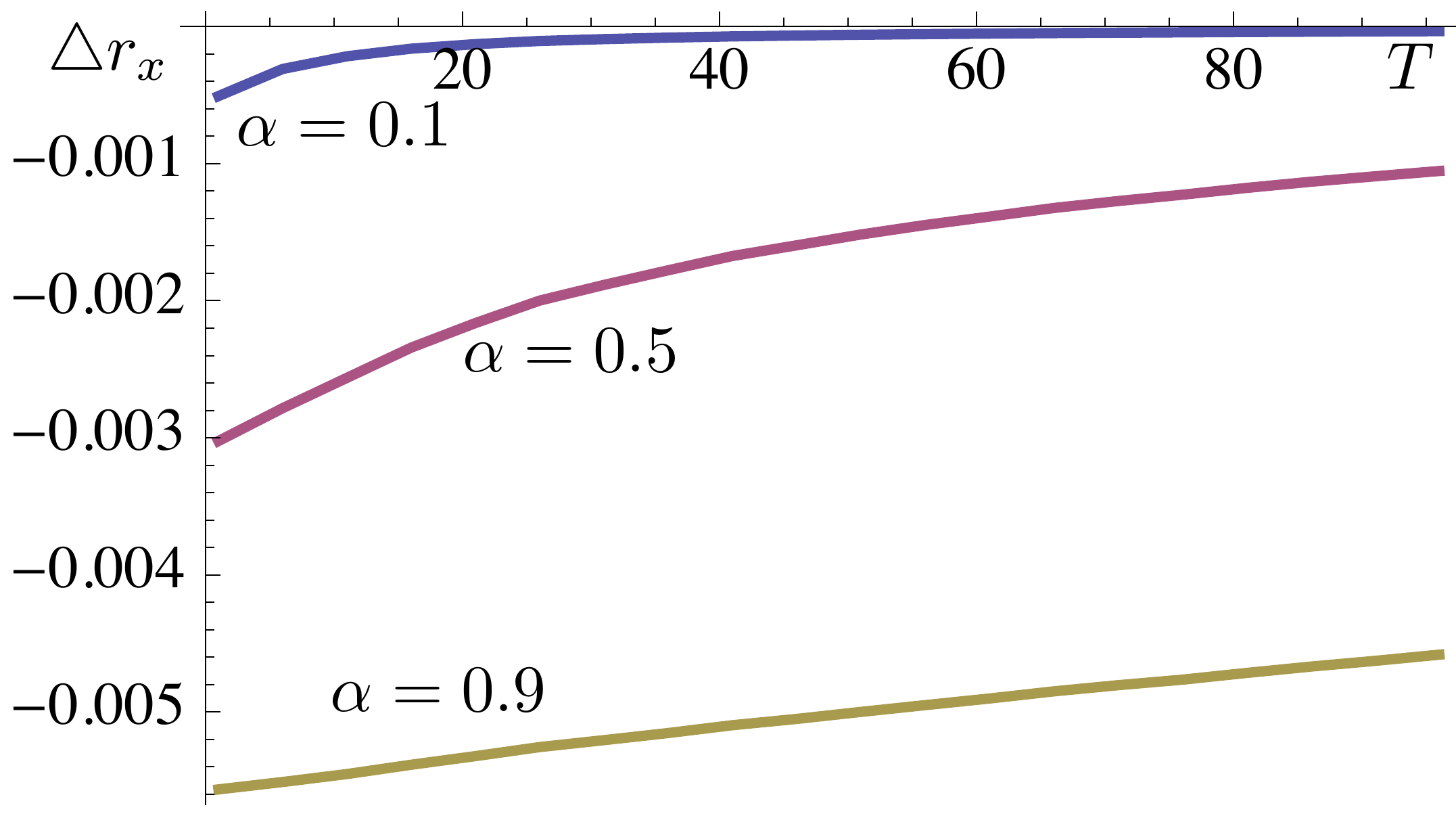}
	\label{fig:slander2}
	}
\caption{\it Effect of the slandering attack. $n=100$.
The difference $\tri r_x$ in the target's reputation is shown (a) as a function of $c$, for
$\vecs=c\vece$; (b) for $\vecs=(1,\cdots,1,0\cdots,0)$ as a function of the number $T$ of pre-trusted users.
(The attacker and target are not part of the pre-trusted users.)}
\label{fig:slander}
\end{figure}

\paragraph{Slandering experiments.}
Again consider one attacker~$y$. His target is $x$.
He sets $A_{xy}=0$, and also makes the following modifications: for $z\notin\{x,y\}$ he sets $A_{zy}=1$ if $A_{xz}<0.5$ and $A_{zy}=0$ otherwise.
(He tries to boost the reputation of those who have a bad opinion about $x$, and to reduce the reputation of the rest.)
The effect is shown in Fig.~\ref{fig:slander}, with the dependence on $\vecs$ presented in the same way as in Fig.~\ref{fig:selfprom}.

We observe that the effect of this attack is roughly ten times stronger than in the self-promotion attack.
The difference  lies in the fact that the attacker $y$ can directly manipulate $A_{xy}$ in the slandering attack, while there is no such possibility in the self-promotion attack ($A_{yy}$ is fixed).
We studied the magnitude of the direct and indirect components of the slandering attack separately. 
The results (not reported here due to the lack of space) show that the direct attack is stronger than the indirect one ($\approx$ ten times).

It is worth noting that the curves in  Fig.~\ref{fig:slander1} are almost flat,
i.e. for $\vecs=c\vece$ the effect of the direct attack on $r_x$ via $A_{xy}$ is almost independent on~$c$.
We suspect (but cannot yet substantiate) that the $c$ largely disappears due to the normalization that is inherently present in the definition of the metric (and which is most clearly visible in step 3 of Algorithm~2) in combination with the fact that all users, including the attacker, are pre-trusted.
In contrast, the curves in Fig.~\ref{fig:slander2} are comparable to the ones for the self-promotion attack  when $y$ is not pre-trusted (i.e., $s_y=0$).

In summary, the countermeasures for mitigating slandering attacks are similar to the one for self-promotion attacks.
However, differently form the self-promotion attack, the inclusion of the attacker in the set of pre-trusted users would make the countermeasure completely ineffective as shown in Fig.~\ref{fig:slander1}.

\begin{figure}[!h]
\centering
\subfigure[]{
\includegraphics[width=58mm]{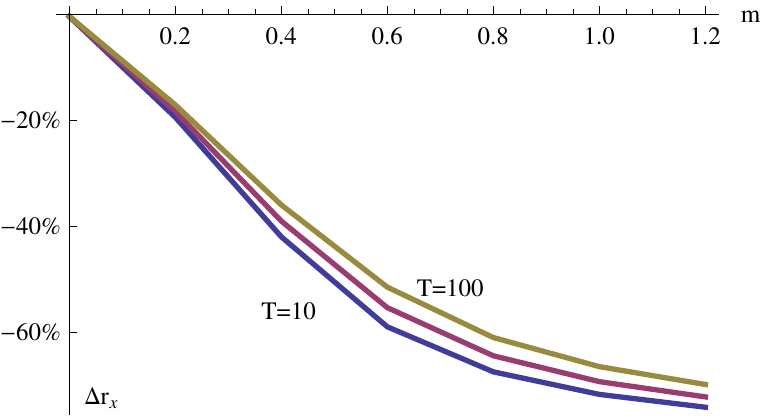}
	\label{fig:sybil1}
	}
\hspace{0.4cm}
\subfigure[]{
\includegraphics[width=58mm]{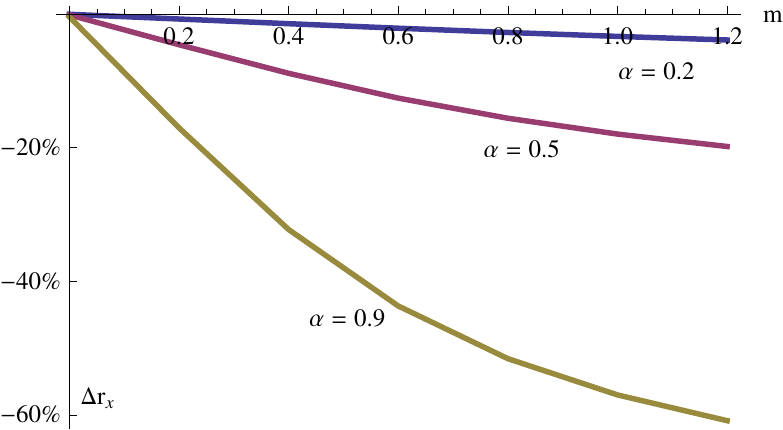}
	\label{fig:sybil2}
	}
\caption{\it Effect of the Sybil attack.
The percentage of the target's reputation reduction is shown as a function of $m$.
In (a) we fixed $\qa=0.9$ and varied the number of pre-trusted users $T$. 
From bottom to top $T=10, 50, 100$. Before the attack, the target's reputation is $0.56$ for all these values of $T$.
In (b) we fixed $T=50$ and varied $\qa$.
(The attacker and sybils are not part of the pre-trusted users; the target is pre-trusted.)
Before the attack, the target's reputation is $0.89$, $0.74$ and $0.53$ for $\qa=0.2, 0.5, 0.9$ respectively.}
\label{fig:sybil}
\end{figure}

\paragraph{Sybil attack experiments.}
We consider one attacker $y$ who creates $n\cdot m$ extra accounts (`siblings')
$n+1,\cdots,n+nm$.
His main aim is to decrease $r_x$ for some fixed target~$x$.
To this end, all the siblings give negative ratings to the target and positive ratings to each other and to $y$.
This corresponds to set $A_{x\qs}=0$, $A_{\qs \qs'}=1$, $A_{y \qs}=1$, where $\qs,\qs'>n$ and $\qs'\neq\qs$.
Furthermore, they gives positive ratings to those users who have rated the target negatively and negative ratings to those users who have rated the target positively.
In our model, this corresponds to set $A_{z\qs}$ to 1 for those $z\neq x,y$ that have $A_{xz}<0.5$, and to $0$ otherwise. 
We started with $n=200$ and in each experiment we increased the size of $A$ 
by adding pseudonyms to the set of users such that the pseudonyms make up between $0\%$ and $120\%$ of the original number of users. 
The effect of the Sybil attack is shown in Fig.~\ref{fig:sybil}.
Clearly the attack is much more effective than the slandering attack in Fig.~\ref{fig:slander}.
As expected, the effect grows with the numbers of siblings. 
Notice that a very large number of siblings is required to significantly reduce $r_x$;
at $m=1$ (as many siblings as original users) still about 40\% of the target's reputation remains.

In Fig.~\ref{fig:sybil1} we can see that increasing the number of pre-trusted users helps to improve the robustness against Sybil attacks; however, the choice of the starting vector $\vecs$ has little effect on the attack.
Fig.~\ref{fig:sybil2} shows that $\qa$ has a nontrivial impact. 
Indeed, with small values of $\qa$ we give less weight to $A$ and, as consequence, the attack is less strong too. Finally, both figures show that the 
effect per added sibling is strongest for $m$ smaller than approximately $0.6$, and for larger $m$ saturation sets in; 
more and more siblings have to be added to obtain significant effect.

%% file: conclusions.tex
\section{Conclusions}
\label{sec:conclusions}

We have presented a flow-based reputation metric for aggregated feedback.
The metric gives absolute reputation values instead of merely a ranking; 
it also makes use of all the relevant information without discarding any part of it, 
leading to reputation values with better discriminating capabilities.
We have given a proof that there is always a solution and that it is unique.
We have also compared different methods for computing the reputation vector, 
and studied the properties of the metric numerically, focusing in particular on how attackers can manipulate reputation values.
We have analyzed the impact of the initial reputation vector $\vecs$ and the weight parameter $\qa$ on $\vecr$.
It turns out that the reputations depend on $\qa$ in a surprisingly linear way, although 
the equations are nonlinear.
They interpolate between the known solutions at $\qa=0$ and $\qa=1$, with small deviations from a straight line.
The direct information plays an important role (also for the ranking) even when little weight is given to it.

We have also studied how these parameters can be used to make the reputation metric more robust against attacks.
The attacks can be direct (attacker $y$ manipulates $A_{xy}$ for target $x$) as well as indirect
(manipulating $A_{zy}$ for other users $z\neq x$). 
A Sybil attack increases the effectiveness. 
The most evident result is that the $\qa$ parameter has a much stronger effect on the robustness than $\vecs$.
Robustness against attacks and in particular Sybil attacks is obtained by choosing a smaller $\qa$.
However, a balance must be kept between resisting attacks and making constructive use of the information provided by users in the $A$ matrix. 
In particular, $\qa$ must not be chosen too small because there is a danger from choosing a wrong $\vecs$.
Setting a larger $\qa$, the effect of choosing the wrong pre-trusted users is mitigated, as the choice of $\vecs$ hardly matters.
This is demonstrated by the jumps in Fig.~\ref{fig:sdep2}, which have size $1-\qa$.

In this paper, we have mainly focused on the mathematical model of the reputation metric.
In particular, we have studied its properties both analytically and numerically, which allows the specification of guidelines for making the system more robust against attacks.
An interesting challenge for future research is to study whether those properties are preserved in the computation dimension, in particular, when there does not exist a centralized authority computing reputation values, but the computation is distributed across the users of the system.

%% file: proofs.tex
\noindent
\underline{Proof of Lemma~\ref{lemma:unit}}\newline
Eq.~(\ref{selfconsistent}) states that $\vecr$ is the weighted average of two
vectors; the first of these vectors is $\vecs\in[0,1]^n$; 
the second vector is $\vecg:=A\vecr/\ell$.
The weights are $\qa$ and $1-\qa$ respectively.
We have $g_a=\sum_b (r_b/\ell)A_{ab}$, i.e. $\vecg$ is the weighted average
of all the columns of~$A$, with weights $r_b/\ell$;
these weights are all nonnegative and add up to~1.
Hence, since $A_{ij}\in[0,1]$, the $\vecg$ satisfies
$g_a\in[0,1]$ for all $a$.
From the fact that both $\vecs$ and $\vecg$ have entries only in $[0,1]$,
it follows that their weighted average has the same property.
\hfill$\square$

\vskip1mm

\noindent
\underline{Proof of Theorem~\ref{th:exist}}\newline
First we prove the existence of a solution.
For $\ell>\qa\qlmax$ Lemma~\ref{lemma:square} tells us that 
$(\one\frac\ell\qa-A)^{-1}\geq0$; hence $f(\ell)\geq 0$.
Furthermore, $f(x)$ is a decreasing function of $x$ on this interval.
Next we use $\qa\qlmax/\ell<1$ to express the matrix inverse as a convergent 
Taylor series,
\be
	\left(\ell\one-\qa A\right)^{-1}=\ell^{-1}\sum_{k=0}^\infty
	\left(\frac\qa\ell\right)^k A^k.
\label{Taylorinv}
\ee
Each term is nonnegative, $(A^k)_{ij}\geq0$.
Next we use the bound $A\leq C$, where $C$ is the 
`constant' matrix. 
It has the special properties $C^k=n^{k-1}C$ (for $k\geq 1$)
and $\veceT C=n\veceT$.
This gives 
$A^k \leq n^{k-1}C$
and allows us to bound the inverse as follows,
\bea
	\left(\ell\one-\qa A\right)^{-1} &\leq& 
	\ell^{-1}\left[\one
	+C\sum_{k=1}^\infty \left(\frac\qa\ell\right)^k n^{k-1}\right]
	\nn\\ &=&
	\ell^{-1}\left[\one+C\frac{\qa/\ell}{1-\qa n/\ell}
	\right].
\eea
Using this bound, and $\veceT\vecs\leq n$ and $\veceT C\vecs\leq n^2$,
we can bound $f(n)$ as
$f(n)\leq 1$.
	
Next we investigate the function $f(\ell)$ in the limit $\ell\downarrow\qa\qlmax$.
The matrix $(\ell\one-\qa A)^{-1}$ has only nonnegative components, and its 
component $(\ell-\qa\qlmax)^{-1}\vecvmax \vecvmax^{\rm T}$ blows up.
We have seen in Section~\ref{sec:alphaspecial} that $\vecvmax>0$. Furthermore, we have $\vecs\geq 0$ and $\vecs\neq{\bf 0}$.
Hence $\vecvmax^{\rm T}\vecs>0$. We conclude that $\lim_{\ell\downarrow\qa\qlmax}f(\ell)=\infty$.

From all the above it follows that $f(\ell)$ on the interval $(\qa\qlmax,n)$ is a decreasing function
spanning at least the whole range $[1,\infty)$, 
and hence has to intersect the value $1$ for some $\ell$. 
This proves the existence of a solution 
$\ell_*\in(\qa\qlmax,n]$ of (\ref{fell}),
which implies that $\vecr=\vecu(\ell_*)$ is a solution of~(\ref{selfconsistent}).

Finally we prove that this solution satisfies $\vecr\in[0,1]^n$.
From Lemma~\ref{lemma:square} we know that $(\one\ell_*/\qa-A)^{-1}\geq 0$.
Substitution into (\ref{uell}) and using $\vecs\geq 0$ gives $\vecr\geq 0$.
From Lemma~\ref{lemma:unit} it then follows that $\vecr\in[0,1]^n$.
\hfill$\square$

{\it Remark:} We have restricted ourselves to irreducible $A$ in Def.~\ref{def:A}.
However, if $A$ is reducible then in almost all cases Theorem~\ref{th:exist} still holds.
For reducible nonnegative $A$ the Perron-Frobenius theorem gives $\vecvmax\geq 0$
instead of $\vecvmax>0$.
The proof above hinges on $\vecvmax^{\rm T}\vecs>0$. 
This condition is satisfied as long as $\vecs$ is not perpendicular to $\vecvmax$.
For instance, if $\vecs>0$ then automatically  $\vecvmax^{\rm T}\vecs>0$.
Furthermore, 
for $\vecs\geq 0$ and randomly generated $A$, the probability of the event $\vecvmax^{\rm T}\vecs=0$ is negligible.

\vskip1mm

\noindent
\underline{Proof of Corollary~\ref{corol:limitsqa}}\newline
In the limit $\qa\to 0$, (\ref{fell}) directly gives $\ell\to\veceT\vecs$
and (\ref{uell}) gives $\vecr\to \vecs$, as expected.
The limit $\qa\to 1$ is less straightforward.
Let us write the decomposition of $\vecs$ into eigenvectors of $A$ as
$\vecs=\sum_i d_i \vecv_i$.
Then, (\ref{fell}) is solved by 
$\ell_*=\qa\qlmax+(1-\qa)\veceT\vecvmax d_{\rm max}$ (which has the correct limit
$\ell_*\to\qlmax$). Substituting $\ell_*$ into (\ref{uell})
precisely yields~(\ref{ralpha1}).
\hfill$\square$

\vskip1mm

\noindent
\underline{Proof of Theorem~\ref{th:unique}}\newline
From the fact that $f(\ell)$ is monotonically decreasing,
it follows that the $\ell_*$ given by Theorem~\ref{th:exist}
is the only solution of $f(\ell)=1$ on the interval $\ell>\qa\qlmax$.
Next we consider solutions $\ell'$ on the interval $(0,\qa\qlmax)$.
In order for $\vecu(\ell')$ to be nonnegative, it has to satisfy
$\veca^{\rm T}\vecu(\ell')\geq 0$ for all $\veca>0$.
If we can find a counter-example then we know that $\vecu(\ell')$ is {\em not} nonnegative.
One counterexample is $\veca=\vecvmax$. From (\ref{uell}) we have
$\vecvmax^{\rm T}\vecu(\ell')=(1-\qa)(\vecvmax^{\rm T}\vecs)/(1-\qa\qlmax/\ell')$,
which is negative for $\ell'<\qa\qlmax$.
\hfill$\square$

\vskip1mm

\noindent
\underline{Proof of Theorem~\ref{th:drdA}}\newline
Eq.~(\ref{selfconsistent}) can be written as 
$[\vecr-(1-\qa)\vecs]\ell = \qa A\vecr$.
The first order part of this equation
(linear in $\qd A$ and $\qd\vecr$)
is given by
$\ell\qd\vecr+[\vecr-(1-\qa)\vecs]\veceT\qd\vecr
= \qa\qd A\vecr+\qa A\qd\vecr$.
Gathering together all the terms multiplying $\qd\vecr$ and $\qd A$, 
then using $\vecr-(1-\qa)\vecs=(\qa/\ell)A\vecr$,
and finally
isolating $\qd\vecr$, we get
\[
	\qd\vecr = \qa\left[ 
	\ell\one-\qa A+\frac\qa\ell A\vecr\veceT
	\right]^{-1}\qd A\;\vecr.
\]
In index notation it reads
\be
	\qd r_x=\sum_{zy}
	\left[\ell-\qa A+\frac\qa\ell A\vecr \veceT\right]^{-1}_{xz}\;(\qd A)_{zy}\; r_y.
\ee
We also know from elementary differential calculus 
that $\qd r_x=\sum_{zy}\frac{\prt r_x}{\prt A_{zy}}(\qd A)_{zy}$.
\hfill$\square$

\vskip1mm

\noindent
\underline{Proof of $\vecr_1=(1+\frac\qa{\ell_0})\vecr_0$}\newline
Let us modify $A$ to $A'=A+\qz\one$, with $\qz\in[0,1]$.
The solution of (\ref{selfconsistent}) using $A'$ will be denoted as $\vecr_\qz$.
Thus we have $\vecr_\qz=(1-\qa)\vecs+\qa A'\vecr_\qz/\veceT\vecr_\qz$.
Next we try if there is a solution of the form $\vecr_\qz = k \vecr_0$, where $k$ is some constant.
This yields $k\vecr_0=(1-\qa)\vecs+\qa(A+\qz\one)\vecr_0/\ell_0$. We use (\ref{selfconsistent})
to replace the expression $(1-\qa)\vecs+\qa A\vecr_0/\ell_0$ by $\vecr_0$.
This yields $(1+\qa\qz/\ell-k)\vecr_0={\bf 0}$.
Since $\vecr_0\neq{\bf 0}$ we conclude that $k=1+\qa\qz/\ell_0$.
Theorem~\ref{th:unique} guarantees that the found solution $\vecr_\qz=k\vecr_0$ is unique.
\hfill$\square$